\documentclass[a4paper,fleqn,usenatbib]{mnras}

\usepackage[T1]{fontenc}
\usepackage{ae,aecompl}

\usepackage{graphicx}	
\usepackage{amsmath}	
\usepackage{amssymb}	
\usepackage{multicol}   

\newcommand{\sigg}{\Sigma_{\text{g}}}

\title[galactic anatomy]{The anatomy of a star-forming galaxy: Pressure-driven regulation of star formation in simulated galaxies}
\author[S. M. Benincasa et al.]
{
Author List}
\author[S. M. Benincasa et al.]
{
S. M. Benincasa\thanks{E-mail: benincsm@mcmaster.ca}, J. Wadsley, H. M. P. Couchman and B. W. Keller\\
Department of Physics and Astronomy, McMaster University, Hamilton, ON L8S 4M1, Canada}

\date{Accepted XXX. Received YYY; in original form ZZZ}

\pubyear{2015}

\begin{document}
\label{firstpage}
\pagerange{\pageref{firstpage}--\pageref{lastpage}}
\maketitle

\begin{abstract}
We explore the regulation of star formation in star-forming galaxies through a suite of high-resolution isolated galaxy simulations.  We use the SPH code \textsc{Gasoline}, including photoelectric heating and metal cooling, which produces a multi-phase interstellar medium.  We show that representative star formation and feedback sub-grid models naturally lead to a weak, sub-linear dependence between the amount of star formation and changes to star formation parameters.  We incorporate these sub-grid models into an equilibrium pressure-driven regulation framework.  We show that the sub-linear scaling arises as a consequence of the non-linear relationship between scale height and the effective pressure generated by stellar feedback. Thus, simulated star-formation regulation is sensitive to how well vertical structure in the ISM is resolved.  Full galaxy disks experience density waves which drive locally time-dependent star formation.  We develop a simple time-dependent, pressure-driven model that reproduces the response extremely well. 
\end{abstract}


\section{Introduction}
The process of star formation is limited, in principle, by the availability of cold, dense gas fuel.  However, in typical disk galaxies it proceeds inefficiently relative to characteristic time-scales for the gas, such as the local free-fall time in Giant Molecular Clouds ($\sim 10$ Myr).  In fact, globally, star formation proceeds on a time-scale comparable to several galactic rotation periods \citep[$\sim$1 Gyr,][]{kennicutt98,krumholzTan07}.  
This is commonly attributed to self-regulation.  
Highly resolved samples of nearby star-forming galaxies, such as The HI Nearby Galaxies Survey \citep[THINGS,][]{THINGSpaper}, the PdBI Arcsecond Whirlpool Survey  \citep[PAWS,][]{PAWSpaper}, and the Panchromatic Hubble Andromeda Treasury \citep[PHAT,][]{PHATpaper} have provided
a detailed look at the interstellar medium (ISM) on sub-kpc scales.  This lets
us connect the local physics of the ISM to the regulation of star formation on larger scales\citep[e.g.][]{krumholzMT09}.

Although observations indicate that star formation must be regulated overall, the details of regulation are difficult to pin down using observational data due to the timing offset between tracers of gas and tracers of star formation \citep[e.g.][]{kruijssenLongmore14}.  Thus the physical cycle of regulation is most easily studied theoretically. Simulations using isolated galaxies have tended to focus on exploring different kinds of feedback and overall star formation rates \citep{hopkinsQM11,hopkinsQM12}.  Galaxy-scale simulations cannot directly resolve the detailed process of regulation, leading to increasingly complex sub-grid models.  Studies of star-formation regulation have also been extended into the cosmological context \citep{agertz13,agertzKravtsov15}.  Going to larger scales includes more of the galactic environment at the cost of lower resolution which potentially compromise elements of the regulation mechanism.

\cite{ostriker10} presented a detailed semi-analytical model of equilibrium star formation and applied it to the THINGS galaxy sample.  In this model, star formation is regulated by satisfying two equilibria in a galaxy.  The first, a vertical dynamical equilibrium, requires a balance between ISM weight and pressure support.  The second, a thermal or energy equilibrium, requires energy balance between feedback and heating/cooling processes \citep{ostriker10,ostrikerShetty11,kimKimOstriker11,kimOstriker15}.   In this work we do not need to make any explicit assumptions regarding energy balance as the simulations do that explicitly.

Following the first requirement, the ISM is compressed by its own weight, set by the gas column and the gravity of the various components of the galaxy (gas, dark matter and stars).  The weight is then a function of the height of the gas layers.  This sets an expected pressure at the mid-plane, which determines the cold gas fraction via the two-phase instability.  Star formation occurs steadily in cold, dense gas.  To keep this model consistent, the ISM must have sufficient means of support to provide that mid-plane pressure.  In the \cite{ostriker10} model, this pressure is effectively set by the star formation rate.  The equilibrium semi-analytic model provided an excellent fit to star formation rates in the THINGS galaxies, where the vertical gravity is dominated by stars.

Determining the total effective pressure is key for vertical pressure balance.  The dominant mechanisms providing this effective pressure support can change in different environments.  In most nearby galaxies ($\Sigma_g \gtrsim 10$ M$_{\odot}$ pc$^{-2}$), however, pressure from turbulent support plays the major role.  This model assumption has been tested on small scales and in two- and three-dimensional simulations \citep{kimKimOstriker11,ostrikerShetty11}.  Stellar feedback is responsible for generating this support.  This may come from supernovae, however, in starbursts, some authors argue that alternate mechanisms, such as radiation pressure, play a major role \citep{hopkinsQM11}.  Supernovae as regulators of star formation have been explored through small-box simulations in several works \citep[e.g.][]{deAvillez05,joung09,creasey13}.

In the prior scenarios, the effective pressure is strongly linked to the local star formation rate.  We note in passing that in the outer regions of disks, ($\Sigma_{\text{g}} \lesssim 10$ M$_{\odot}$ pc$^{-2}$), gas could be supported due to non-local heating or turbulence associated with galactic shear \citep{mcnally09}.  A two-phase structure may not develop, limiting star formation in these environments \citep{elmegreenParravano94, schaye04}.  

Local simulations with a fixed surface density are conducive to reaching a steady equilibrium and thus ideal to study the hydrostatic, equilibrium framework of the \cite{ostriker10} model.  Pressure-driven regulation should also play a key role in disk-scale simulations, with the addition of time dependent, spatial variations.  Isolated galaxy simulations naturally include all of the necessary components (e.g. large-scale shear) but still allow relatively high resolution for the ISM.  High-resolution simulations of isolated galaxies have been extensively used to study the formation of dense star-forming gas \citep{dobbs08,dobbs11,renaud13,taskerTan09,tasker11}.  However, a realistic feedback and star-formation regulation cycle was not a major focus of these works.

Examining pressure balance and star-formation regulation requires high
numerical resolution.  Fundamentally, one must be able to resolve the scale height of the gas disk, particularly for the colder phases, where stars form, which have scale heights of order 100 pc or less.  Forming multiple phases is also numerically demanding.  Turbulence is an important contributor to the detailed structure of star forming clouds.  As a result, galaxy-scale simulations can directly examine star cluster-scale formation at best.  Thus even high-resolution simulations must still rely on star formation recipes.  Turbulence also contributes to regulation and pressure support.  Galactic turbulence is generated at a range of scales up to several kpc and then cascades down to smaller scales.  Simulations
have difficulty maintaining this cascade.  In particular turbulent energy is typically suppressed (via numerical dissipation) on scales substantially above the resolution limit \citep{priceFederrath10}.  In galaxy-scale simulations this severely limits turbulence on small scales in the ISM.  However, turbulence on scales comparable to the scale height is strongly linked to stellar feedback and thus star-formation regulation.  An attractive option is to avoid the numerical issues by injecting energy on small scales as an effective pressure as in \cite{agertz13}.

The pressure-driven regulation framework has profound implications for
simulators.  The fact that most simulated models are able to regulate star formation suggests that just providing a source of effective pressure linked to young stars is sufficient for basic regulation.   In addition, star formation and feedback models typically have a lot of parameter freedom to tune star formation rates to match expectations.  Thus achieving regulation or even tuning it to match a narrow set of observations is not that remarkable.  The pressure-driven picture should explain these results if we incorporate typical sub-grid models into the framework.

For example, a common outcome from the prior work discussed above has been that simulated star formation rates do not scale linearly with variations in star formation and feedback parameters \citep[e.g.][]{hopkinsQM11}.  Linear scaling in the star formation rate constant, for example, would be the naive expectation.  This assumes that star formation and the ISM are weakly coupled.   However, variations in key ISM properties such as scale height are likely to be artificially limited at low resolution.  By moving to higher resolution than has typically been employed for many sub-grid models in use today, one can probe how robust these results are.

The above issues with sub-grid star formation in whole-galaxy simulations undermine the common approach of using regulated star formation as the primary benchmark for a successful model.   The key question raised is whether the regulation achieved in simulated galaxies is actually similar to that occurring in nature.  An associated question is how much predictive power these models have when used outside the cases on which they were calibrated.

The pressure regulation idea is extremely general and should apply to all steadily evolving galaxies, even as conditions and modes of feedback are varied.  Small scale simulations have confirmed the utility of these ideas for local, relatively uniform conditions \citep[e.g.][]{kimKimOstriker11, kimOstrikerKim13}.  We would like to apply this insight to understand simulations on the scale of a whole galaxy.   To do so we should extend the framework to include the time dependent behaviour, (e.g. in response to density waves), expected in the broader galactic context.   To do so, one needs to move from vertical hydrostatic equilibrium to a non-equilibrium pressure-driven framework.

In the current work, we employ a suite of high resolution simulations of isolated galaxies to explore basic ideas of star-formation regulation.  Though we use high resolution and incorporate a fairly complex ISM model, we have kept the star formation and feedback prescriptions simple to make the results easier to interpret.  None-the-less, these models are quite similar to the most popular star formation and feedback models in use in simulations at the current time and should therefore also provide direct insight into how those models operate.

Given that our resolution is much higher than much of the early work in which the sub-grid models were developed \citep[e.g.][]{stinson06}, a first step is to test basic sensitivity to parameter choices.  We can also compare to other recent, high-resolution simulations of isolated galaxies, which have tended to use much more complex feedback \citep[e.g.][]{hopkinsQM11}.  

A further feature of this work is to test the pressure-driven framework in a dynamic environment with time dependence and spatial variations (e.g. density waves) arising naturally in a galactic setting.  This necessitates extending the model into a time-dependent, (non-equilibrium) pressure-driven framework.

The structure of the paper is as follows.  In section~\ref{sec:method}, we describe our simulation set-up for an isolated galaxy and our simple feedback recipe.  In section~\ref{sec:results}, we characterize the overall simulated galaxy behaviour.  In particular, it shows sub-linear scaling of star formation rates with respect to the model parameters and we explore how these arise.   In section~\ref{sec:balance}, we demonstrate that the galaxies were behaving in a manner consistent with expectations from the pressure balance framework.  Section~\ref{sec:timevary} demonstrates the local time variability in star formation and associated ISM properties.  Finally, in section~\ref{sec:dynamicpressure}, we present a time-dependent model extension of the pressure-driven framework.  This reproduces the behaviour seen in the simulations including the sub-linear scaling with star formation parameters.

\begin{figure}
\centering
\includegraphics[scale=0.42]{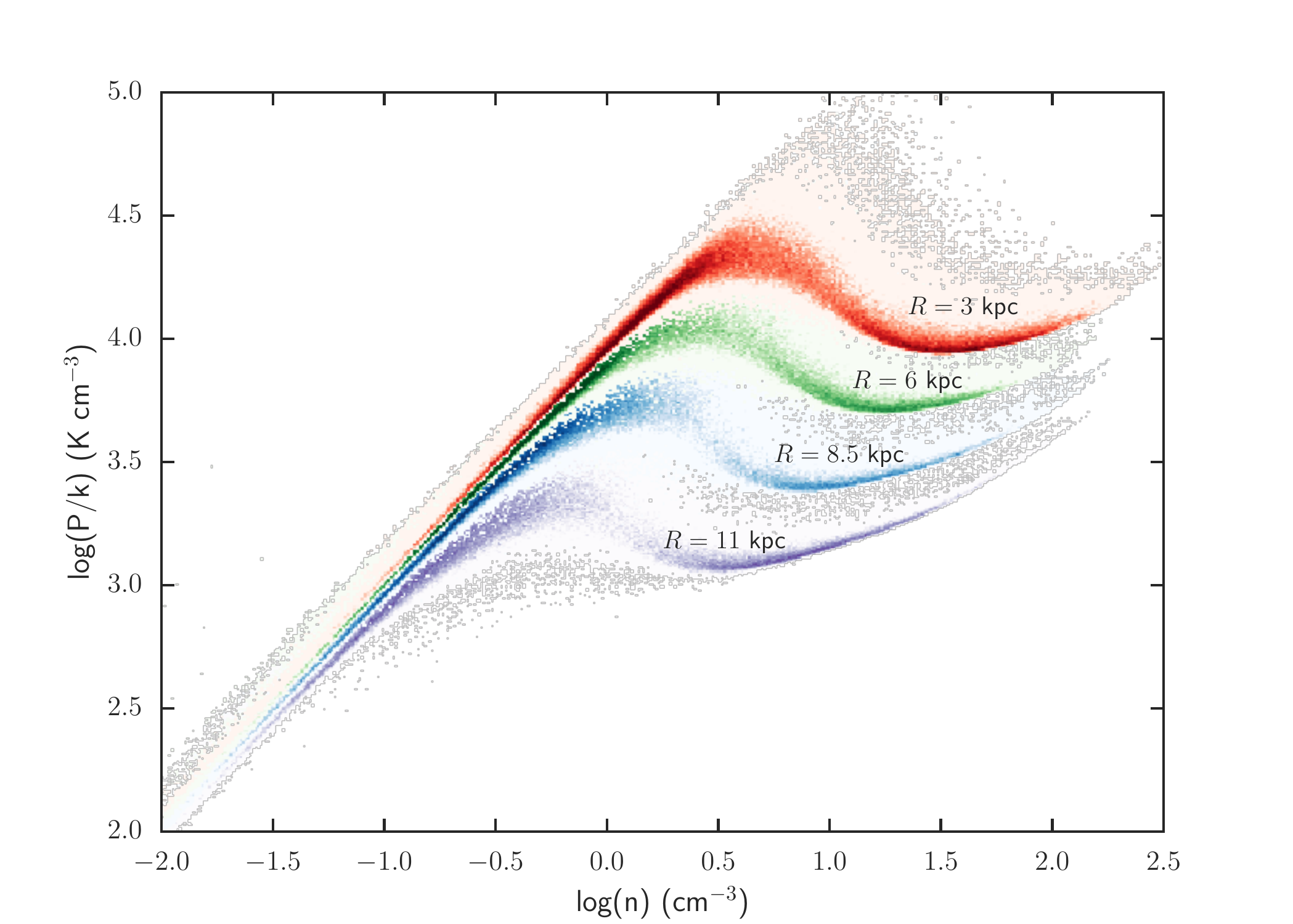}
\caption{Phase diagram for the reference galaxy at a time of 500 Myr.  Gas was sampled in annuli centred at the labelled radii +/- 50 pc, with different colouring denoting different radii.  Here, darker colouring denotes more mass residing at that location in the phase space.  Since the photoelectric heating term varies with radius, the equilibrium is different at different radii in the disk.  This variation changes the minimum density required to maintain a two-phase structure in the ISM.}
\label{fig:phase}
\end{figure}

\section{Simulation Method}
\label{sec:method}
We use the modern SPH code \textsc{Gasoline} \citep{wadsley04} including recent improvements \citep[with details described in][]{keller14}.  Our simulations are high resolution, with a particle mass of 442 M$_{\odot}$ (force softening length of 20 pc).   These choices allow us to resolve the scale height of the disk.  Our simulations also form GMC-scale dense clouds, meaning we can maintain two-phase
structure in our ISM \citep[for a detailed study of GMCs formed in our simulations see][]{wardBenincasa16}.

This resolution is similar to that used by \cite{hopkinsQM11}.  The new features used in this work include a treatment for photoelectric heating in the ISM and a simpler feedback model, described below.

\subsection{Photoelectric heating}
\label{ssec:peheating}
The dominant source of heating due to photoelectric absorption by dust grains comes from Ultra-Violet (UV) radiation from stars.  The heating rate due to photoelectric heating is set by
\begin{equation}
n \Gamma_{\text{PE}} = 10^{-24}\epsilon n G_0 \text{ erg cm}^{-3} \text{ s}^{-1}
\end{equation}
where $\epsilon$ represents a heating efficiency, $G_0$ is the intensity of the radiation field in units of the average interstellar radiation field and $n$ is the number density of the gas.  In this work we assume that $\epsilon G_0 \sim 0.05$ \citep{tielensBook}.  To model this photoelectric heating throughout the disk we employ a FUV heating term similar to \cite{taskerTan09}:  
\begin{equation}
n \Gamma_{\text{FUV}} = n  \Gamma_{\text{PE}} \times \left\{
\begin{array}{l c l}
 e^{-(4-R_0)/4} & : & R < 4 \ \text{kpc}\\
 e^{-(R-R_0)/4} & : & R \ge 4 \ \text{kpc}\\
\end{array}
\right.
\end{equation}
with $R_0=8.5$ kpc, where $R$ is a cylindrical radius. This functional form is chosen to match the FUV profile used
in \cite{wolfire2003} which was derived from an assumed distribution of young
stars consistent with the Milky Way.  Thus, for the current work the FUV heating
is not directly coupled to simulated star formation events.  However, the
typical Galactic optical depth to FUV is fairly low, with mean free paths of
order a kpc. A smoothed FUV field is a reasonable approximation and this was assumed
by \cite{wolfire2003} in deriving a radial profile.  We expect star formation to
occur throughout the disk in a distribution similar to the one that gives rise
to this FUV field and this was the outcome seen in our simulation runs.  This
approach has been used in other simulations \citep{dobbs11,tasker11}.  A higher
degree of consistency would require simulated radiative transfer.

We also approximate the depletion of elements onto dust grains. This is done by multiplying the total metal abundance by a constant
depletion factor of 0.4, similar to that assumed in \cite{wolfire2003}.  This
generates a linear decrease in the metal cooling which gives a closer match to
the heating and cooling behaviour in \cite{wolfire2003}.

The outcome of this treatment for photoelectric heating is shown in Figure~\ref{fig:phase}.  The phase diagram is shown for our reference galaxy at the
final output time of 500 Myr.  The phase diagram is plotted for four different
radii; 3, 8.5 and 11 kpc (for comparison, see \cite{wolfire2003} Figure 7).  For
the four tracks shown, areas of dark colour correspond to areas in the phase
space containing more mass.  In each case we see a clear preference for gas to
be in either a warm or cold ISM phase.  Our simulations are successful at
producing an ISM with two-phase structure.  Note the finite thickness and
presence of some material in the unstable regime.  These effects are due to the
dynamics present in the simulations which occur on time-scales similar to or
smaller than the characteristic time required for complete two-phase separation.

\subsection{Star formation and feedback}
\label{ssec:fb}
The star formation recipe used here is as described in \cite{stinson06} with some modifications. In the recipe, stars form following a Schmidt law:
\begin{equation}
\frac{\text{d}\rho_*}{\text{d}t} = c_* \frac{\rho_g}{t_{dyn}},
\end{equation}
where $\rho_*$ is the density of new stars formed, $\rho_g$ is the density of eligible gas, $t_{dyn}=1/\sqrt{4\pi G \rho}$ is the dynamical time and $c_*$ is the chosen efficiency (see section~\ref{sec:results}).  Gas is considered eligible for star formation if it lies above a set density threshold, below a maximum temperature and belongs to a converging flow.

The original recipe required that gas particles be below a set temperature threshold to form stars.   We now compare the effective temperature associated with the total energy (including thermal and non-cooling, as discussed below) to the threshold value.  This ensures that feedback will always locally limit star formation.  In this work the temperature threshold value is taken at 100 K.  The key star formation model parameters that apply here are the minimum density threshold for gas to form stars ($n_{\text{th}}$), the star formation efficiency per dynamical time ($c_*$) and the feedback efficiency ($\epsilon_{\text{fb}}$) relative to the standard amount from supernovae, as described below.  

We employ the supernova feedback method of \cite{agertz13}.  This model differs from the older blastwave model of \cite{stinson06} and the newer superbubble model of \cite{keller14} in that it is a simpler way to deposit the energy from SNe that still produces similar results and robust feedback regulation.  Specifically, rather than being a complex function of local properties, the conversion time for injected energy is a set parameter.  This gives us direct knowledge of how long feedback energy remains available as effective pressure and removes the associated uncertainty in interpreting our results.  In more detailed models, feedback energy might take the form of pockets of hot gas, cosmic rays, radiation pressure, winds and small-scale turbulence, all injected on initially very small scales \citep[$\sim$pc,;see e.g.][]{hopkinsQM12}.  Many of these processes are difficult to model, convert energy rapidly from one form to another and suffer from numerical effects including excess dissipation.  The associated effective pressures become important for driving dynamics when they act on scales comparable to the scale height.  A simple model ensures a well characterized coupling of feedback energy to the dynamics on these scales.  It also allows us to side-step the unresolved issue of the relative importance of different types of feedback energy and how efficiently it converts into different forms.

\begin{figure*}
\centering
\includegraphics[scale=0.75]{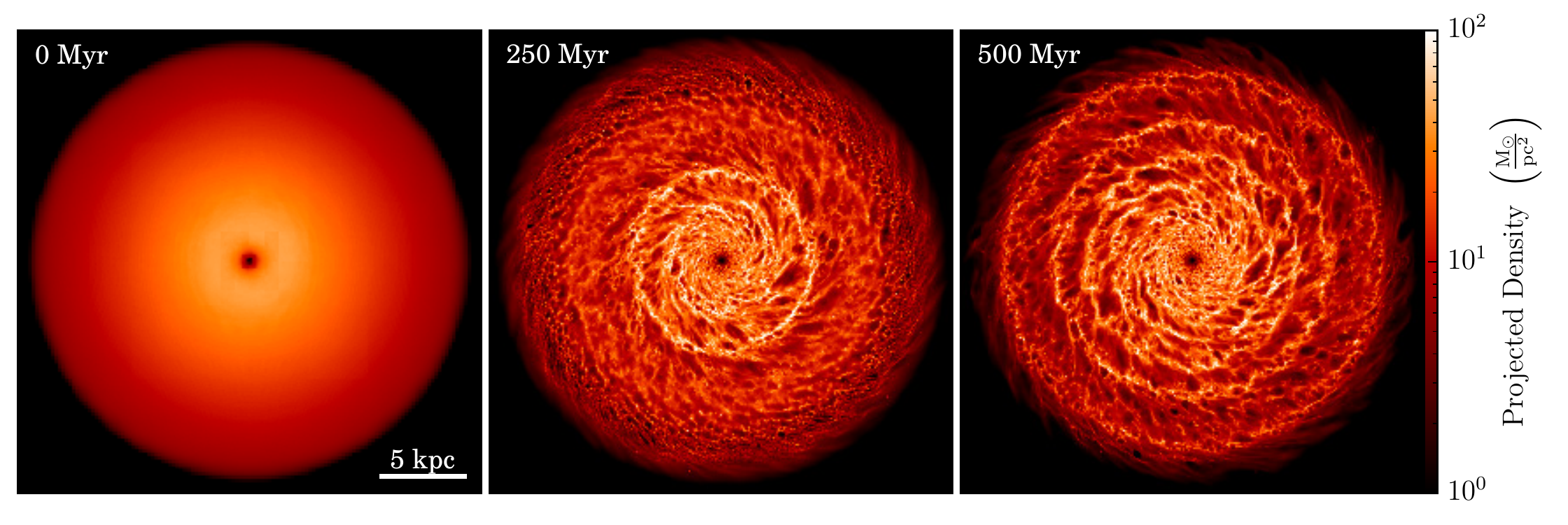}
\caption{Three surface density maps for the reference galaxy (n100.c6) at different times.  \textit{Left:} a snapshot of the initial condition from which each simulation is evolved.  \textit{Middle:} a snapshot at 250 Myr, halfway through the evolution of the galaxy.  \textit{Right:} a snapshot at 500 Myr, the point at which each simulation is stopped.  The ring structures discussed in the text are clearly visible as the disk evolves.  Each snapshot is 30 kpc across.}
\label{fig:evolution}
\end{figure*}

We normalize our energy input relative to that of supernovae.  A single supernova
deposits $10^{51}$ erg of energy into the surrounding ISM.  Based on the stellar
initial mass function of \cite{chabrierIMF}, approximately one in 100 stars will
undergo a supernova event, resulting in an average specific supernova energy of $10^{49}$
erg/M$_{\odot}$.  On average, this energy is injected steadily over the first 40
Myr after the formation of a star cluster.  In this method, feedback energy,
$E_{\text{fb}}$, does not cool radiatively and is steadily converted into regular thermal energy at a fixed rate similar to \cite{agertz13}.
The relevant energy equations are,
\begin{subequations}
 \begin{align}
  \frac{\text{d}u_{\text{th}}}{\text{d}t} & = \frac{u_{\text{fb}}}{\tau} + \dot{u}_{\text{th},P\text{d}V} - \Lambda\\[0.5cm]
  \frac{\text{d}u_{\text{fb}}}{\text{d}t} & = -\frac{u_{\text{fb}}}{\tau} + \dot{u}_{\text{fb},P\text{d}V} + \dot{u}_{\text{fb},*} \label{eqn:unc}
 \end{align}
\end{subequations}
where $u_{\text{th}}$ is gas thermal energy, $u_{\text{fb}}$ is the non-cooling energy, $\dot{u}_{P\text{d}V}$ represents energy exchanges due to hydrodynamics, $\Lambda$ is the cooling rate and $\dot{u}_{\text{fb},*}$ is new feedback energy injected by stars.  Feedback energy can be decreased through PdV work pushing the gas around and is also steadily converted back into the regular, cooling form.  This latter change occurs over a chosen conversion time-scale, $\tau$.  For the purpose of this work, we adopt a conversion time of $\tau=5$ Myr.  This energy deposition due to supernovae will begin immediately after a star cluster forms.  This covers the period where other forms of feedback are effective such as stellar winds.  Thus in this model, the exact nature of the feedback is not specified, just that the total energy is measured relative to the expected supernova amount and that it is injected over a time period from the formation of the star cluster for a period of 40 Myr as determined by the supernova rate.

\subsection{Initial conditions}
\label{ssec:IC}
We generated our suite of high resolution galaxy simulations around a common galaxy model.  Detailed descriptions of parameter choices for each run are discussed in section \ref{ssec:suite}   The isolated disk galaxy model was given a density profile decreasing exponentially with radius as follows,
\begin{equation}
 \Sigma_g(R) = \Sigma_0 \left(\frac{R^2}{R^2 + 1}\right) e^{-(R-8)/5}
\end{equation}
where $R$ is a cyclindrical radius in kpc and $\Sigma_0=$14 M$_{\odot}$ pc$^{-2}$, chosen to model a Milky Way-type galaxy.  This results in a disk with a total gas mass of 7.51$\times 10^9$ M$_{\odot}$.  For comparison, the gas mass of the Milky Way is approximately $4.5\times 10^9$ M$_{\odot}$ \citep{tielensBook}.  We exclude the centre of the disk at radii interior to 1 kpc.  By excluding this region, we do not have to resolve nuclear star formation. Choosing to have an unresolved galaxy centre, or not have a centre at all are common strategies \citep[e.g.][]{dobbs08, benincasaTasker13}.

The disks were evolved in a static dark matter halo potential.  A log halo potential was used:
\begin{equation}
 \Phi_\text{L} = \frac{1}{2}v_0^2 \text{ln}\left(R_c^2 + R^2 + \frac{z^2}{q_{\Phi}^2}\right) 
\end{equation}
where $R, \ R_c$ and $z$ are in units of kpc.  This gives a circular speed as a function of radius of:
\begin{equation}
 v_c = \frac{v_0 R}{\sqrt{R_c^2 + R^2}}
\end{equation}
\citep{binneyTremaine}.  For our galaxy, $q_{\Phi}=1$, $R_c = 1$ kpc, and $v_0 = 220$ km $\text{s}^{-1}$.  Again, we choose $v_c$ to match the rotation velocity at the solar radius in the Milky Way.

\section{Simulations}
\label{sec:results}

\subsection{The simulation suite}
\label{ssec:suite}
The suite of high resolution isolated galaxy simulations primarily explored the three parameters from the star formation and feedback 
recipes: the density threshold ($n_{\text{th}}$), the star formation efficiency
per dynamical time ($c_*$) and the feedback efficiency ($\epsilon_{\text{fb}}$).
We chose as our reference parameter set a density threshold of
$n_{\text{th}}=100$ cm$^{-3}$, a star formation efficiency of $c_*=6\%$ and a
feedback efficiency of $\epsilon_{\text{fb}}=100\%$.  Our choice of threshold
coincides with the average density of GMCs.  Our reference efficiency per
dynamical time of 6\%, corresponds to $\sim$3\% per free-fall time ($t_{ff}=\sqrt{3\pi /32 G \rho}$).
For comparison, an efficiency of 5\% is a common choice in galaxy simulations.
We also chose efficiencies to match those of \cite{hopkinsQM11}
for ease of comparison.

The full set of parameter choices for each simulation in the suite is listed in
Table~\ref{tab:params}.  The simulations are named following the convention
n($n_{\text{th}}$).c($c_*$).(comments).  The feedback efficiency is listed in
the comments placeholder only if it deviates from our reference choice of 100\%.
For instance, our reference simulation with $n_{\text{th}}=100$ cm$^{-3}$,
$c_*=6\%$ and $\epsilon_{\text{fb}}=100\%$ is labelled n100.c6.

\begin{table}
\footnotesize
 \begin{center}
 \caption{List of Simulations \label{table:sims}}
 \begin{tabular}{lccc}
& & & \\
 \hline
 \hline \\[-2ex]
  name & $n_{th}$ (cm$^{-3}$)&$c_*$ & $\epsilon_{\text{fb}}$\\
 \hline \\[-2ex]
 n100.c25 & 100 & 25\% & 100\%\\
n100.c6$^\dagger$ & 100 & 6\% & 100\%\\
n100.c6.FB30 & 100 & 6\% & 30\%\\
n100.c6.FB10 & 100 & 6\% & 10\%\\
n100.c1.5 & 100 & 1.5\% & 100\%\\
n100.c0.35 & 100 & 0.35\% & 100\%\\
n300.c6 & 300 & 6\% & 100\%\\
n1000.c6 & 1000 & 6\% & 100\%\\
 \hline
 \multicolumn{4}{l}{$^\dagger$ The reference parameter set.}\\
 \label{tab:params}
 \end{tabular}
 \end{center} 
\end{table}

\begin{figure*}
\centering
\includegraphics[scale=0.63]{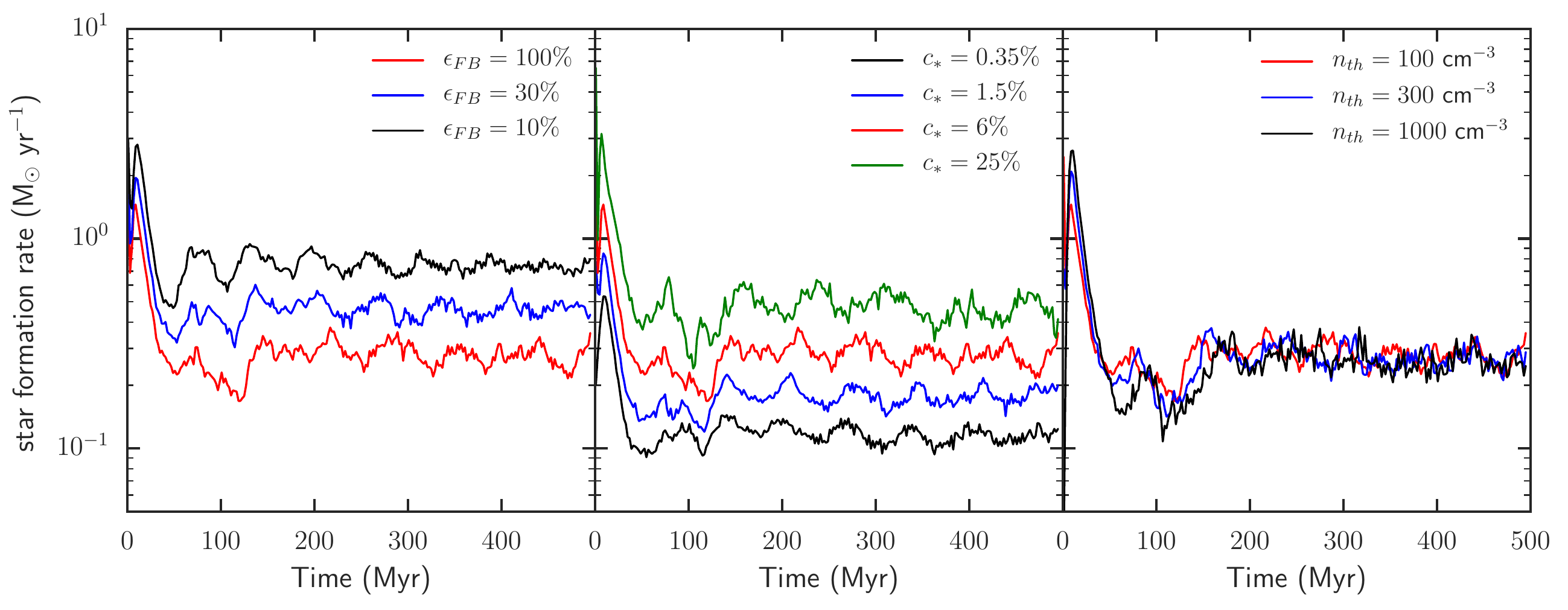}
\caption{The global star formation rates for our suite of galaxies.  For each panel, the red line always denotes our fiducial galaxy.  Left: the result of changing the feedback efficiency ($\epsilon_{fb}$), holding all else constant.  Middle: the result of changing the star formation efficiency ($c_*$).  Right: the result of changing the density threshold ($n_{th}$).  Raising the density threshold shows no accompanying change in the SFR, while raising either star formation or feedback efficiency causes sub-linear changes in the SFR (see text).}
\label{fig:sfr}
\end{figure*}

Figure~\ref{fig:evolution} shows a density projection of our fiducial galaxy at
three different times.  By 500 Myr we see strong density ring features.
As mentioned above, our model includes no old stellar disk.  These rings are a
side effect of the lack of gravity from this excluded component which would
impose a dominant spiral potential if present.  All of the galaxies in the
suite exhibit these rings to a varying extent.  Increasing or
decreasing $c_*$ increases or decreases the strength of the ring structures,
respectively.  This is similar to the behaviour for changes in feedback
efficiency.  When the efficiency of SNe is decreased to 10\% (n100.c6.FB10), the
prominent rings disappear.

\subsection{The star formation rate}
\label{ssec:sfr}
We begin by comparing the global Star Formation Rates (SFRs) for each simulation
in the suite in Figure~\ref{fig:sfr}.  Overall
the rates are at the low end compared to nearby galaxies \citep[e.g.][]{bigiel08}, even taking into account the quiet central region.   We attribute this to the
absence of an old stellar disk which causes a substantial reduction in the vertical gravity which would have decreased the scale height and increased the star
formation rate.  More specifically, the star
formation models of \cite{ostriker10}, as applied to THINGS disk galaxies, are
heavily influenced by the old stellar disk and most new star formation occurs
there.  We defer examination of the role of an older stellar disk and associated spiral modes to future work.  For now, this simpler picture is easier to interpret and compare to other isolated simulations of self-regulation.

There are a number of important features in the plots.   First, is the scaling between the SFRs and changes in simulation parameters. The SFR is insensitive to a
change in the threshold density required to form a star; changing the threshold
by an order of magnitude does nothing to the global SFR.  These findings are consistent with those of \cite{hopkinsQM11}, where changes in star formation parameters do not change SFRs in the Milky Way-type galaxy.  Changes in threshold parameters are difficult to interpret in a global model as the density threshold is a parameter that impacts star formation on GMC scales.

The SFR is sensitive to efficiency-type parameters,
specifically, those governing the star formation efficiency per dynamical time
and the amount of feedback energy used.  This sensitivity is not linear in the
efficiency in either case, however.  With respect to $c_*$, a change from 0.35\%
to 25\%, a factor of 64 in efficiency, changes the star formation rate by a
factor of 5 on average.  In general, a change in $c_*$ by a factor $f_*$,
results in a change in the SFR by a factor of approximately $f_*^{0.3}$.  With
respect to $\epsilon_{\mathrm{fb}}$, a change from 10\% to 100\%, results in an increase in SFR by a factor of 2.5; in general a change in $f_{\mathrm{fb}}$, results in a change in the SFR by a
factor of approximately $f_{\mathrm{fb}}^{0.4}$.  All changes in the star
formation rate are found to scale \textit{sub-linearly} with changes to
parameters.  To better understand the origins of this scaling, we must consider
the behaviour of the ISM.

\begin{figure*}
 \centering
  \includegraphics[scale=0.75]{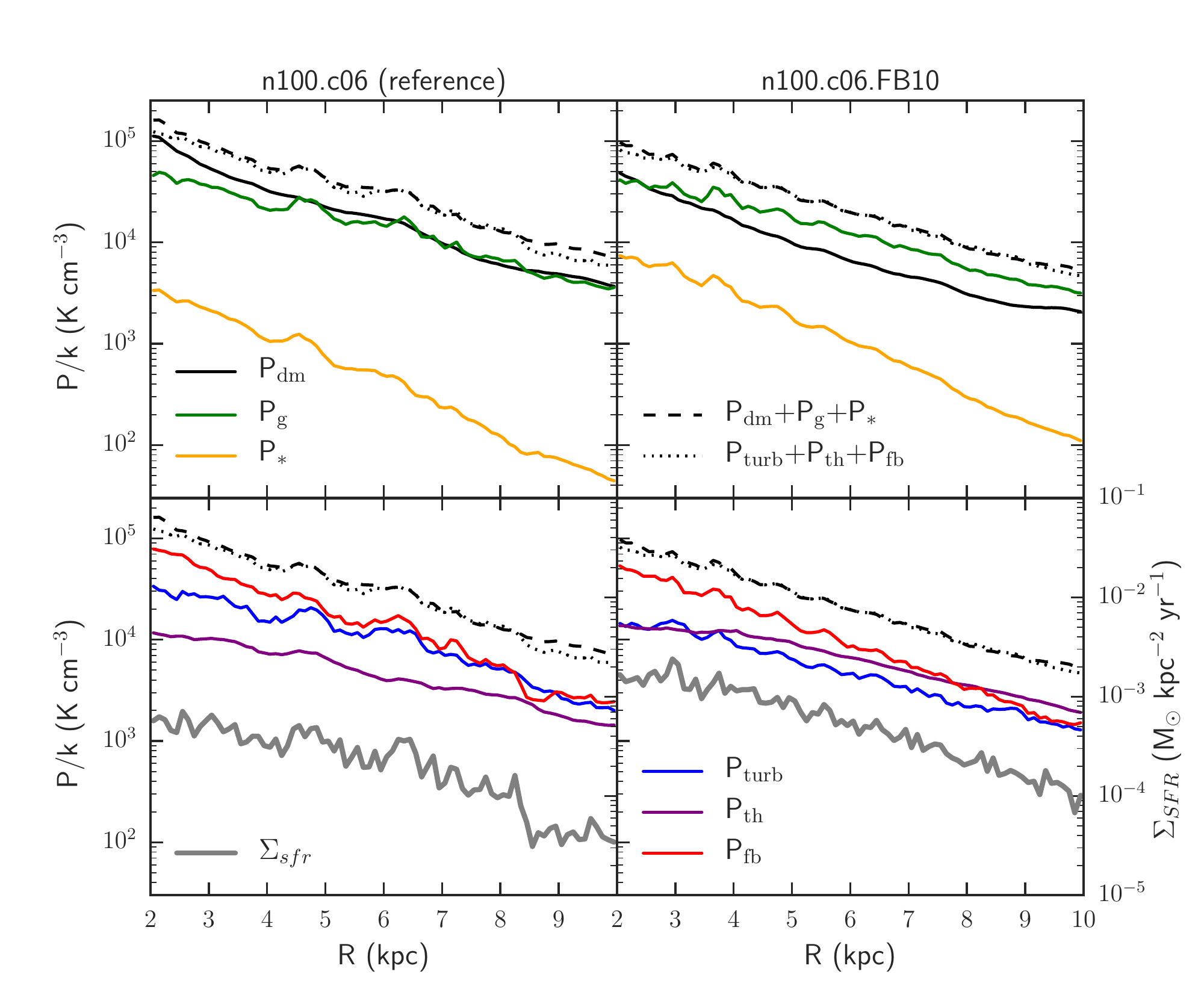}
 \caption{\textit{Left:} average effective pressures for the reference simulation. \textit{Right:} average effective pressures for the low feedback efficiency simulation (see Section~\ref{ssec:suite} for details).  The top panels show the effective pressure required to balance vertical gravity from the three components of the disk; dark matter (black), gas (green) and stars (orange).  The bottom panels show the provided effective pressure support from turbulence (blue), thermal energy (purple) and feedback energy (red).  The total pressure required to support vertical gravity (dashed line) is balanced by these effective pressure sources (dotted line).  For comparison we also plot the SFR surface density (thick grey line, righthand scale). Each line shows an average over 100 Myr.}
 \label{fig:pFiducial}
\end{figure*}

Another characteristic feature is the oscillation in the rates.  Specifically, although each of
the rates settles to a fairly well-defined average, there is ongoing periodic
variation about this equilibrium.  Further, the frequency of this oscillation
appears to be similar for all cases. This corresponding time-scale is
approximately 75 Myr.  This can be compared to the various time-scales important
for the disk material. Firstly, there is the vertical oscillation time of the
galaxy; in the Milky Way at the solar radius this time is 84 Myr
\citep{binneyTremaine}.  Due to the lack of old stars and associated surface density, the vertical oscillation time is longer at 220 Myr.  Second,
there is the time-scale over which star clusters deliver feedback.  This
feedback delivery time can cause bursts or ring structures in the disk at times
when star formation itself is experiencing a burst.  This occurs on a time-scale
of approximately 40 Myr.

\section{What holds up a galaxy?}
\label{sec:balance}

In order to get an idea for what sets the star formation rate for a given galaxy, and better understand the scaling between this rate and other quantities we examine this behaviour within a pressure-driven framework for the regulation of star formation.  We begin with equilibrium behaviour which is very similar to the vertical dynamical equilibrium portion of the framework developed by \cite{ostriker10} and \cite{ostrikerShetty11}.

In the original pressure-driven framework, gas weight associated with vertical
gravity is exactly matched by effective pressure support expected when star formation proceeds at a steady rate.  Considering the weight of all disk components we arrive at an expression for the effective pressure \textit{required} to maintain equilibrium:
\begin{align}
 \mathrm{P}_{\mathrm{R}} &= \mathrm{P}_{dm} + \mathrm{P}_g + \mathrm{P}_* \nonumber \\ 
&= \frac{1}{2}\Omega^2\Sigma_g \mathrm{H}_g  +  \frac{1}{2}\pi G \Sigma_{\text{g}}^2 + \pi G\Sigma_g\Sigma_* \label{eqn:ptotcomp}
\end{align}
where $\Omega$ is the shear rate, $\Sigma_g$ is the gas surface density, $\mathrm{H}_g$ is the gas scale height and $\Sigma_*$ is the stellar surface density.  Detailed assumptions used in deriving these terms can be found in appendix~\ref{app:pressuremath}.
\begin{figure*}
\centering
\includegraphics[scale=0.75]{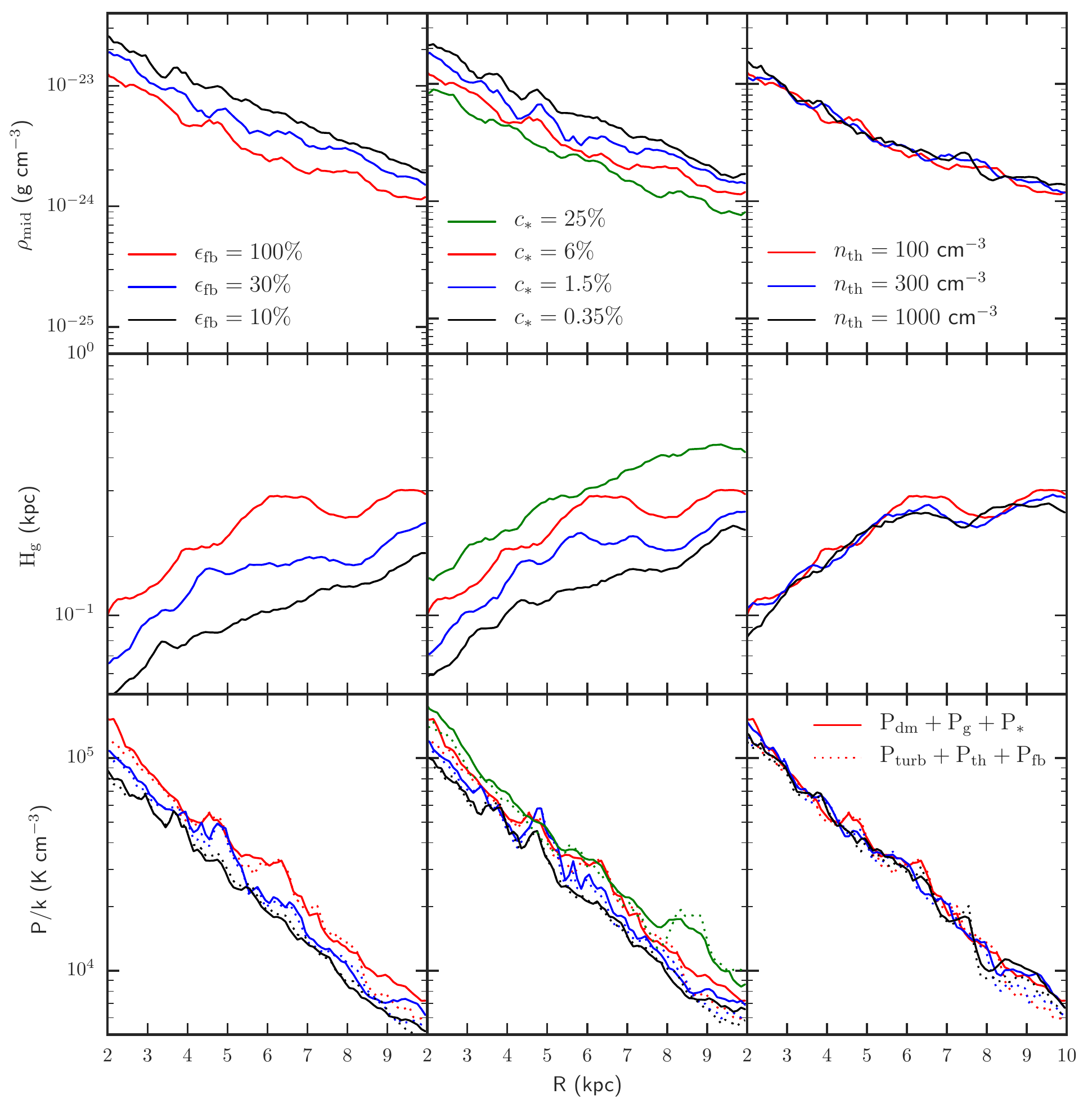}
\caption{Summary of the three main galaxy regulation properties of interest.  Top row: the effective mid-plane density.  Middle row: the scale height of the gas disk.  Bottom row: the effective pressure.  In this row dashed lines represent the total pressure support and solid lines represent the total pressure required. }
\label{fig:main}
\end{figure*}

Equation~\ref{eqn:ptotcomp} indicates the level of pressure support required for
the gas to avoid vertical collapse.  In the galaxy support can come from sources which are linked to or independent of local star formation.  Turbulence provides effective pressure
where the original energy can come from galactic shear or stellar feedback
processes.  Thermal pressure, originating from UV radiation, photoelectric heating and dissipation of gas motions, can provide support.  Sources of non-thermal
pressure, such as magnetic fields and cosmic rays could also contribute.  However, the scale height of cosmic rays is larger than that of the gas disk and so should not be a major factor.  As noted above, in a simulation this translates to pressure support that is linked to local star formation (e.g. stellar feedback) or independent of local star formation (e.g. magnetic fields, cosmic rays, etc...).  For the purposes of this study
we consider only the support provided by turbulent or thermal pressure sources explicitly.

For small-scale contributions associated with stellar feedback, we do
not specify the exact form that the pressure takes but refer to it simply as
effective pressure associated with feedback, $\mathrm{P}_{fb}$.  For the purpose of
providing support, the form is not important.  Using our feedback model, it ultimately translates into thermal
energy, $u_{th}$, on a fixed time-scale of 5 Myr as in Equation~\ref{eqn:unc} .  It can also push around gas, losing
energy through PdV work and potentially increasing the effective turbulent support associated with vertical motions,  $v_z$.  In this work the pressure \textit{support} we measure in the disk is calculated by:

\begin{align}
  \mathrm{P}_{\mathrm{S}} &= \mathrm{P}_{th} + \mathrm{P}_{fb} + \mathrm{P}_{turb} \nonumber \\
    &= \frac{\Sigma_{\mathrm{g}}}{2\,H_g} \left(
\frac{2}{3}u_{\mathrm{th}} + \frac{2}{3}u_{\mathrm{fb}} + v_z^2
\right)_{z=0}. \label{eqn:ptotsupport}
\end{align}
This is the mid-plane support and all the quantities take on their
mid-plane values.  The mid-plane density, $\rho_{\mathrm{g},0}$, is well
approximated by the gas surface density divided by twice the gas scale height,
$H_g$.
\subsection{Equilibrium behaviour in our simulated galaxies}

Figure~\ref{fig:pFiducial} shows the pressure contributions present for the
reference galaxy in our suite (see section~\ref{ssec:suite}).  The top panel in
Figure~\ref{fig:pFiducial} shows the effective pressure required to balance
vertical gravity from dark matter (black), gas (green) and stars (orange).  The
bottom panel shows the pressure support provided by resolved turbulence (blue), thermal
(purple) and feedback (red) energy.  In both panels the dashed line is the total
pressure as calculated by equation~\ref{eqn:ptotcomp} and the dotted line is the
total pressure as calculated by considering thermal and turbulent effective
pressures (as in equation~\ref{eqn:ptotsupport}).  Finally, the SFR surface density for each case is plotted in the bottom panels of Figure \ref{fig:pFiducial} (thick grey line). To facilitate a more direct comparison with the equilibrium models, we have smoothed in time, averaging over 100 Myr (from an output time of 400 to 500 Myr, to avoid early transients).

Referring again to Figure~\ref{fig:pFiducial}, the two lines denoting required
and supplied support pressure in the disk agree in each case. This reflects the
fact that on average the disk is in pressure equilibrium; star formation is regulated to balance vertical gravity requirements when averaged over long time periods.  Similar agreement was seen in the simulations of \cite{kimKimOstriker11}. 

Due to the lack of an old stellar disk, the total
stellar gravity contribution is minimal in our simulations.
The young stellar population has a small scale height and the expression above is an upper-bound for this case.
As noted previously, this is expected to make the effective pressure in our galaxy (and other similar simulations) lower than in typical nearby disk galaxies.

As mentioned above, we can use the time-averaged state of the galaxy to explore the equilibrium of the disk and how this can impact other galaxy properties.   Since the pressure support in the disk is provided as a consequence of star formation, this must set the star formation rate.  This equilibrium then must also be responsible for the sub-linear scaling between the star formation rate and star formation parameters.

In simulations, parameters are our tools to explore different types of galaxies.  Different parameter combinations can cause different types of required pressure or pressure support to be dominant.  When the feedback energy is high (100\%), the scale height of the galaxy is much larger.  This leads to a larger role for the vertical gravity from the dark matter component (see equation~\ref{eqn:Pdm}).  In the top left panel of Figure~\ref{fig:pFiducial} we see that at small radii the vertical gravity from dark matter does dominate the required pressure.  At large radii there is an similar contribution from both dark matter and gas.  Since the feedback energy is higher we also expect a large amount of the support to come from our feedback pressure ($\mathrm{P}_{\mathrm{fb}}$). We see this in the bottom left panel of Figure~\ref{fig:pFiducial}.  The pressure support provided by feedback is dominant in the inner regions of the disk, matched by support from turbulence in the outer regions of the disk.

Lowering the feedback energy (10\%) leads to a change in the dynamics of the disk.  Although the amount of star formation has increased, the total energy injected by those stars has decreased.  The scale height decreases in response to the decrease in available energy and the gas component becomes the dominant source of vertical gravity (see eqn.~\ref{eqn:Pgas}).  We also see that thermal energy plays a much larger role in setting the amount of pressure support although, at most radii, the feedback energy is still the dominant support term.

Although not shown here, similar trends exist when considering the star formation efficiency, $c_*$.  Lowering the star formation efficiency creates similar results to lowering the feedback efficiency, both result in lowered energy injection rates when compared to the reference parameter set.  This tells us that the injection rate of feedback energy is critical in setting the balance between pressure terms.  A low star formation efficiency (e.g. 0.35\% as in n100.c0.35) will produce the same behaviour in pressure terms that we see when lowering the feedback efficiency.

A key assumption in the \cite{ostriker10} model is the strong correlation between thermal pressure and the total effective pressure.  In the \cite{ostriker10} model the contribution from FUV heating is important primarily because it sets the thermal pressure; the overall pressure (e.g. due contributions such as turbulence) is proportional to that with a multiplier of $\alpha=P_{tot}/P_{th} \sim 5$.  This correlation has been confirmed directly in numerical simulations by \cite{kimKimOstriker11,kimOstrikerKim13} and \cite{kimOstriker15}.  Our results are consistent with this sub-dominant role for thermal pressure, as seen in Figure \ref{fig:pFiducial}.  Detailed studies of this partition would require high resolution local simulations \citep[e.g.][]{kimOstriker15}.

A consequence of this strong correlation is the relationship between the SFR surface density and the total effective pressure.  As seen in Figure \ref{fig:pFiducial} the SFR surface density changes in response to changes in the total effective pressure; the correlation is evident.

\begin{figure}
\centering
\includegraphics[scale=0.55]{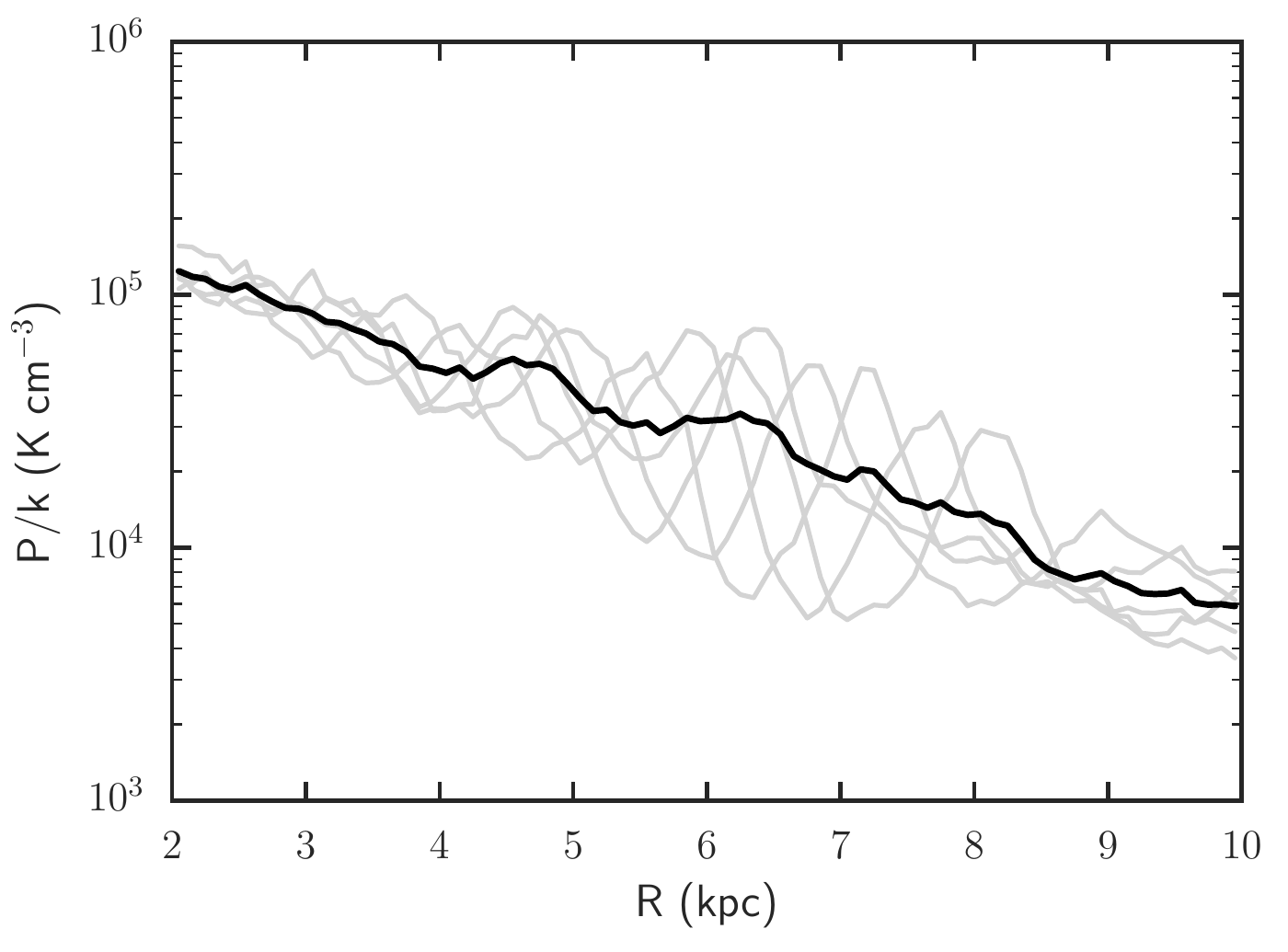}
 \caption{Variation of total pressure at different times for the reference galaxy (n100.c6).  The grey lines show the total pressure, as calculated using equation \ref{eqn:ptotsupport} at 25 Myr intervals between 400 and 500 Myr.  The black line shows the average of the five times.}
\label{fig:timevary}
\end{figure}

Figure~\ref{fig:main} shows a sample of time-averaged galaxy properties.  In all cases, the curves show quantities averaged over a span of 100 Myr (from a time of 400 to 500 Myr, see Figure~\ref{fig:timevary}).  This is the same time span used to produce the average pressure quantities in Figure~\ref{fig:pFiducial}.  The average properties change in response to changes in parameters and here we see similar behaviour to what we have already found in the global SFRs (see Figure~\ref{fig:sfr}).

What we can take away from this figure is that changes that impact global ISM properties are important to global regulation.  For instance, we have previously seen that changes in density threshold do not result in changes to the SFR.  In Figure~\ref{fig:main} there are no changes to the mid-plane density, scale height or pressure as result of the change in threshold.  The density threshold is a parameter that regulates star formation on the scale of GMCs, and so here does not play a role in changing global regulation.  This is essentially the same result found by \cite{hopkinsQM11}.

Parameters which change the available energy \textit{do} impact the global behaviour of the ISM.  As shown, an increase in the feedback efficiency causes a corresponding increase in the scale height.  A larger scale height increases the amount of pressure required.  These together act to decrease the amount of star formation (as seen in Figure~\ref{fig:sfr}).

\section{Time Variation}
\label{sec:timevary}

\begin{figure*}
 \includegraphics[scale=0.85]{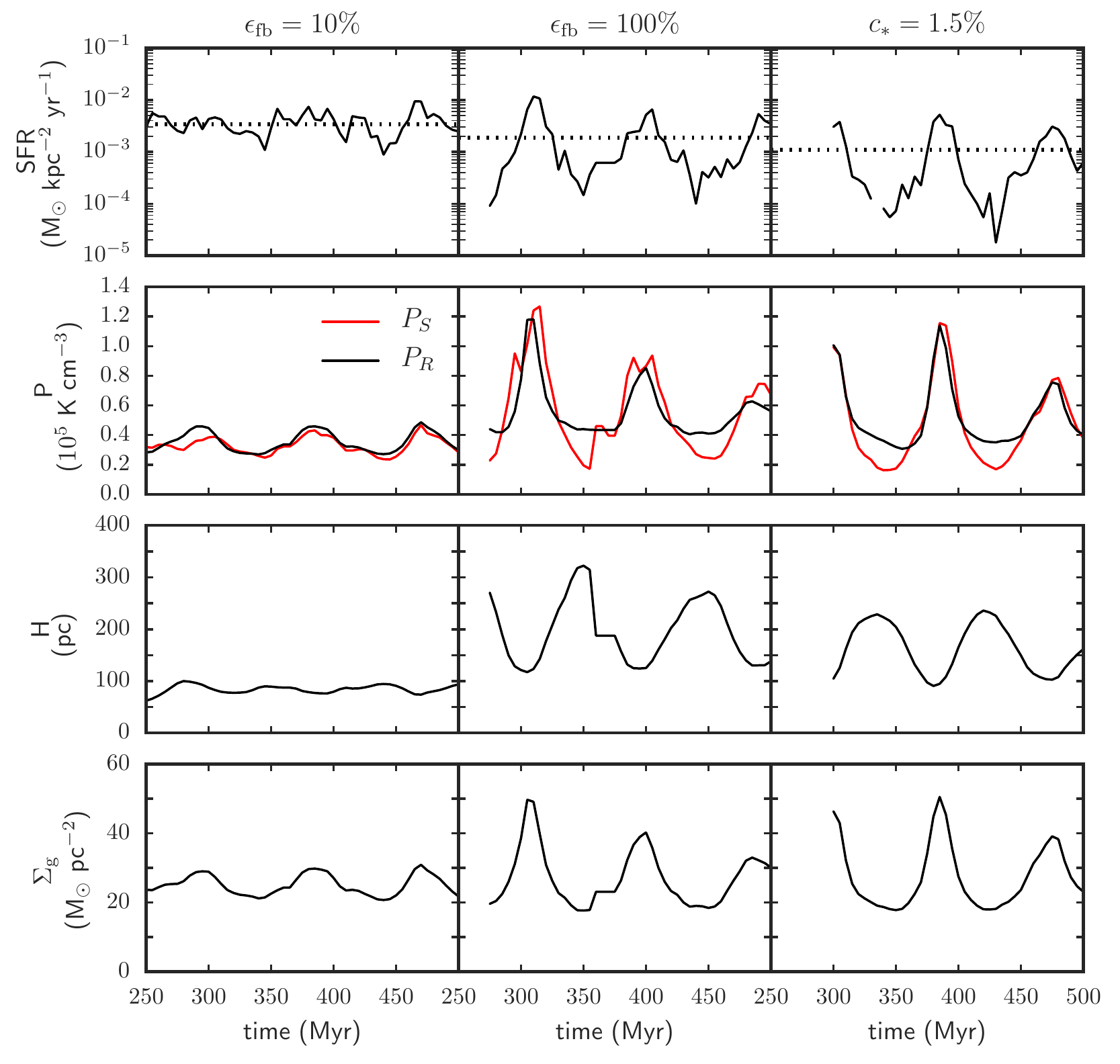}
 \caption{Local behaviour in an annulus of width 100 pc centered on a galactic radius of 4.55 kpc.  \textit{Left:} the behaviour of the ISM for the lowest feedback efficiency simulation (n100.c6.FB10). \textit{Middle}: the behaviour for the reference galaxy (n100.c6). \textit{Right:}  the behaviour for a simulation with a lowered star formation efficiency (n100.c1.5). For a high feedback efficiency, the star formation rate shows clear bursts, this translates to large variations in other quantities.  In contrast, for a lower feedback efficiency, a modest star formation rate can be maintained and this translates to a much smaller range of variation for other ISM quantities.}
 \label{fig:response}
\end{figure*}

\begin{figure*}
\centering
\includegraphics[scale=0.85]{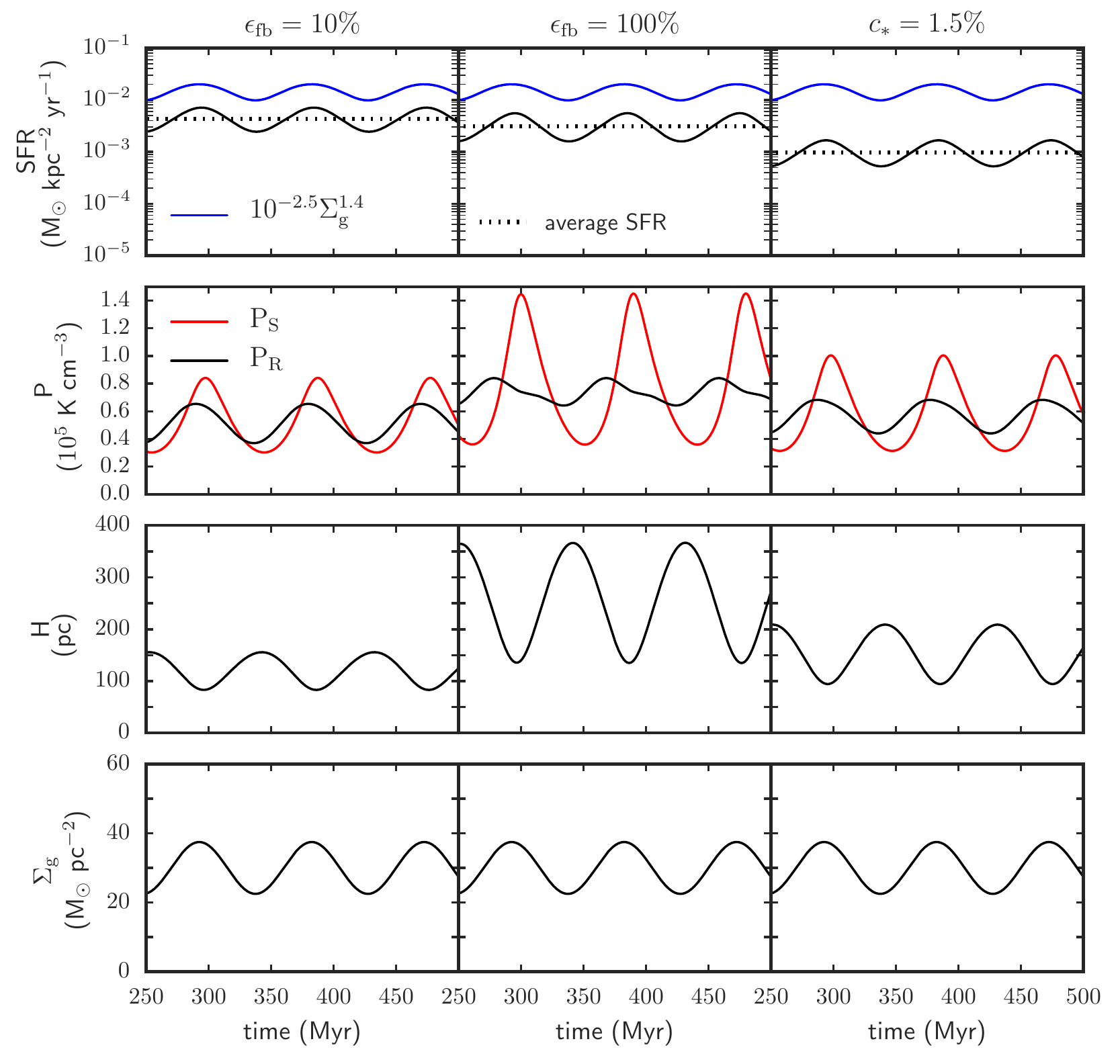}
\caption{Results from our dynamic pressure driven model.  \textit{Right:} a low feedback efficiency case, comparable to our simulation n100.c6.FB10. \textit{Middle:} a high feedback efficiency case, comparable to our reference simulation n100.c6.  \textit{Left:} a low star formation efficiency case, comparable to our simulation n100.c1.5.  Star formation rates are plotted in the top panels, where the black line shows the model rate, the blue line shows what is expected considering the Kennicutt-Schmidt relation \citep{kennicutt98} and the dotted line denotes the average SFR. }
\label{fig:toymodel}
\end{figure*}

\begin{table*}
\footnotesize
 \begin{center}
 \caption{Comparison of the average SFR in annuli at $R=4.5$ kpc for the simulation and our model}
 \begin{tabular}{lccc}
& & \\
 \hline
 \hline \\[-2ex]
 \multicolumn{4}{c}{varying $\epsilon_{\text{fb}}$} \\[1ex]
  name & $\epsilon_{fb}$ &$\overline{\text{SFR}}$ (M$_{\odot}$ kpc$^{-2}$ yr$^{-1}$)&model $\overline{\text{SFR}}$ (M$_{\odot}$ kpc$^{-2}$ yr$^{-1}$)\\
 \hline \\[-2ex]
 n100.c6$^\dagger$ & 100\% & 0.0018 &  0.0031\\
 n100.c6.FB30 & 30\% & 0.0026 & 0.0038\\
 n100.c6.FB10 & 10\% & 0.0034 &  0.0044 \\
 \hline
 \multicolumn{4}{c}{varying $c_*$} \\[1ex]
 name & $c_*$ & $\overline{\text{SFR}}$ (M$_{\odot}$ kpc$^{-2}$ yr$^{-1}$) & model $\overline{\text{SFR}}$ (M$_{\odot}$ kpc$^{-2}$ yr$^{-1}$)\\
 \hline
 n100.c25 & 25\% & 0.0021 &  0.0097 \\
 n100.c6$^\dagger$ & 6\% & 0.0018 & 0.0031\\
 n100.c1.5 & 1.5\% & 0.0011 & 0.00098\\
 n100.c0.35 & 0.35\% & 0.00052 & 0.00027\\
 \hline
 \multicolumn{3}{l}{$^\dagger$ The reference parameter set.}\\
 \label{tab:sfravg}
 \end{tabular}
 \end{center} 
\end{table*}

In the previous section we discussed our galaxies in a hydrostatic equilibrium framework.  We now explore the time variability in our simulations.  Figure~\ref{fig:timevary} shows an example of the variability of the total pressure support for our reference galaxy (n100.c6).  Plotted is the total pressure
($P_{\text{th}}+P_{\text{fb}}+P_{\text{turb}}$) in radial bins.  The grey lines show the pressure at
specific timing points in 25 Myr intervals between 400 and 500 Myr, while the black line shows the
average of these five times (as plotted in Figure~\ref{fig:pFiducial}).  There is large variation in the five lines at
various radii; disturbances move outward in the disk as time proceeds.

We now examine the time dependent response of a galaxy to changes in the star formation rate parameters.  With all else fixed, the parameters cause instantaneous changes to the star formation rate.  However, over time they also change the ISM.  This response leads to net regulation.  As noted in the introduction, star formation parameters represent tuneable knobs which can lead to different
galactic structure and star formation behaviour. 
  Ideally, they would be set through an understanding of the star
  formation process and the related constraint of producing a realistic ISM.  Simulations typically only marginally resolve the ISM so this cannot be used as a constraint.  Instead, gross properties such as the resulting star formation rate are used.  The simulations examined here have sufficient resolution to probe key aspects of the ISM response.

In Figure~\ref{fig:response} we plot three samples of local
behaviour in the disk, one for our reference galaxy (n100.c6), one for our
lowest feedback case (n100.c6.FB10) and one for a lower star formation efficiency (n100.c1.5).  In these plots we have tracked a 100 pc
wide annulus of the disk centred on a galactic radius of 4.5 kpc.  Shown are the
star formation rate (top), pressure support (second), pressure requirement from
gravity (second), scale height (third) and gas surface density (bottom) for the
three simulations mentioned above.  We choose a radius of 4.5 kpc as substantial star formation occurs there and it is representative of the bulk of the star formation ongoing in the disk.

The first thing to note in these plots is that the behaviour of all of the
quantities is periodic.  For example, the characteristic period of the
oscillations in the surface density is $\sim$ 90 Myr.  This oscillation period
is dependent on the galactic radius where the measurement is made.  This period
is seen in all three cases, regardless of the parameter choices. 

Assuming pressure is driving the regulation of star formation, we expect a dynamic simulation to differ from the equilibrium \cite{ostriker10} picture. In the non-equilibrium case there will be lags in time before these responses are realized.  In Figure~\ref{fig:response} we can see the differing response times between star formation and the chosen quantities.  These response times can be confirmed both visually and through time signal analysis.  In the following we refer specifically to the reference simulation (middle panels of Figure~\ref{fig:response}).  Star formation is maximized at or just after a local peak in the mid-plane density or surface density when dense clouds are formed.  The time between a peak in the local surface density and a local peak in star formation is $\sim$5 Myr (or less).  Given that star formation is inefficient on the time-scales of these oscillations, the galaxy's main mechanism to decrease the star formation rate is to increase the scale height.  An increase in the scale height implies that the mid-plane density of the gas is, on average, lower.  This should lead to a decrease in the amount of star formation.  As seen in Figure~\ref{fig:response}, the time after a local burst in star formation before the corresponding local maximum in the scale height is between 40 and 50 Myr: the galaxy cannot respond immediately to a change in the star formation rate.

It is more difficult to pick out similar response times in the lower feedback
case.  Decreasing the amount of available feedback energy changes the mode of
star formation in the disk.  In the middle panel of Figure~\ref{fig:response}, the
star formation is very bursty; the events occur and then in between these the
rate plummets effectively to zero.  In contrast, when the feedback energy is
lowered star formation can be sustained at a significant level for a longer
period of time.  In the top left panel, it is more difficult to pick out a
clear periodic behaviour in the star formation rate, although it does still
exist in the surface density.  This corresponds to a smaller range of variation in
the other ISM quantities as well.  For example, when the feedback efficiency is
high the scale height varies by approximately 200 pc between the minima and
maxima of its cycle.  When the feedback efficiency is lowered we see a variation
of only approximately 50 pc.

 Plotted in the right-most panel of Figure~\ref{fig:response} is the result for a lowered star formation efficiency.  Now, star formation events are more isolated in time and their intensity is decreased.  Lowering the feedback efficiency is comparable to lowering the star formation efficiency.  In each case, we have lowered the injection rate of feedback energy, just in different ways. 

The dotted lines in Figure~\ref{fig:response} denote the average SFR over the given time period.  Here again we see similar sub-linear scaling between the average SFR and changes in parameters.  In fact sub-linear scaling holds again for all quantities discussed.  With this in mind we can now outline a general non-equilibrium framework to model the responses we see in our simulations.

\section{A Dynamic Pressure-Driven Framework}
\label{sec:dynamicpressure}

The scale height of the gas may be defined as,
\begin{align}
  H_g &= \frac{2}{\sigg} {\int^{\infty}_0 \rho_g z\ \text{d}z}.
\end{align}
The scale height changes in response to density waves moving in the plane of the disk as derived in appendix~\ref{app:presmodel}.  An important property of waves is the lack of net motion which makes advection terms less important.  This is precisely true for linear waves and in the appendix  we show it holds for the disks simulated here.

This allows us to develop a local model of the galaxy.  Using equation~\ref{eqn:dhdt} from the appendix and neglecting the advection terms we have, 
\begin{align}
\frac{\partial H_g}{\partial t} = \bar{v}_z.
\end{align}
Thus the evolution of $H_g$ is dominated by the mean vertical velocity in the half-plane,
\begin{align}
    \bar{v}_z = \frac{2}{\sigg}  \int^{\infty}_0 \rho_g v_z\ \text{d}z.
\end{align}
In appendix~\ref{app:presmodel} we show that the mean vertical velocity responds to differences between the mid-plane pressure and the total pressure required by gravity.  Neglecting advection in equation \ref{dvzdt} gives,
\begin{align}
  \frac{\partial \bar{v}_z}{\partial t} &= \frac{2}{\sigg}  \left( \mathrm{P}_{\mathrm{S}}-\mathrm{P}_{\mathrm{R}} \right).
  \label{eqn:pdvzdt}
\end{align}
Thus the change in the mean vertical velocity depends on the difference between vertically integrated weight, which in section~\ref{sec:balance} we called the required pressure, $\mathrm{P}_{\mathrm{R}}$, and the mid-plane supporting pressure, which we have denoted as, $\mathrm{P}_{\mathrm{S}}$.  These equations describe the dynamic pressure balance model.
This model can incorporate the finite response time of the gas and inputs such as feedback that are distributed in time.  We investigate this later in the Section.

To study the basic behaviour, we substitute in expressions for the pressures (i.e. eqn~\ref{eqn:ptotcomp} and eqn.~\ref{eqn:ptotsupport}) We then get a time-dependent equation for the scale height, 

\begin{align}
 \frac{\partial^2 H_g}{\partial t^2 } 
& = \frac{2}{\sigg}  \left( \mathrm{P}_{\mathrm{S}}-\mathrm{P}_{\mathrm{R}} \right) \nonumber \\
& =  \frac{1}{H_g} (\gamma-1) u_{\mathrm{eff}} 
- \Omega^2 H_g  - \pi G \sigg -  \frac{2 \pi G\Sigma_* H_g}{H_g+H_*}, \label{eqn:dh2dt2}
\end{align}
where we have bundled the pressure support terms in a single effective internal energy, $u_{\mathrm{eff}}$ with an effective $\gamma$.  When $\mathrm{P}_{\mathrm{S}} = \mathrm{P}_{\mathrm{R}}$ we recover the equilibrium solution.  Thus, when the left-hand side is zero, we have an equation for the mean scale height, $\overline{H}_g$.  We can examine oscillations about the mean using a small perturbation in height, $\Delta H_g  = H_g - \overline{H}_g$, and retaining terms of order $\Delta H_g$,
\begin{align}
 \frac{\partial^2 \Delta H_g}{\partial t^2 } 
&= -\left (\frac{1}{\gamma} \frac{c_{\mathrm{eff}}^2}{\overline{H}_g^2} + \Omega^2 + \frac{2 \pi G\Sigma_* H_*}{(\overline{H}_g+H_*)^2}\right)\,\Delta H_g,
\end{align}
where we have rewritten $u_{\mathrm{eff}}$ in terms of the associated effective sound speed, $c_{\mathrm{eff}}$.

Because increasing scale height simultaneously lowers the mid-plane pressure while increasing the vertical gravity, the opposing pressure terms act together to stiffen the restoring force relative to that experienced by collisionless components.  The characteristic gas vertical oscillation frequency, $\Omega_g$,  is thus given by
\begin{align}
  \Omega_g^2 = \frac{1}{\gamma} \frac{c_{\mathrm{eff}}^2}{{\overline{H}_g^2}} + \Omega^2 + \frac{2 \pi G\Sigma_* H_*}{(\overline{H}_g+H_*)^2},
  \label{eqn:osc}
\end{align}
and is indirectly dependent on the gas surface density, $\Sigma_g$, through the mean scale height.  Even when $\Sigma_g \ll \Sigma_*$, galactic gas disks tend to evolve to states where self-gravity is relevant so the mean height varies strongly with $\Sigma_g$.  One would expect the oscillation frequency to fall between the collisionless vertical oscillation frequency and the effective vertical crossing frequency.  However, because the terms in eqn. \ref{eqn:osc} all become additive, the frequency is lower than both of those expected.

In present-day spiral galaxies, the vertical gravity near the mid-plane is dominated by the old stellar population.  This increases the collisionless  vertical oscillation frequency without otherwise qualitatively changing the behaviour as the vertical gravity associated with the dark matter and the stellar disk have a similar form, roughly linear with height, $z$.  The effective equation of states leaves $\gamma \sim 1$ for cases with modest feedback.  For example, at 4.5 kpc in our simulations as shown in Figure~\ref{fig:response}, the collisionless vertical oscillation period, $\frac{2\,\pi}{\Omega}$ = 130 Myr, whereas the time-scale for vertical gas oscillations can be as low as $\frac{2\,\pi}{\Omega_g}$ = 30-60 Myr depending on the mean support.

The gas is not oscillating freely, but is responding to the drivers of changes in the surface density.  The period of the density waves is largely independent of the vertical response of the gas.  In our simulations, the oscillations in surface density occur on characteristic time-scales of 90 Myr, as shown in Figure~\ref{fig:response} and also in the time variation in the star formation rates in Figure~\ref{fig:sfr}.  Thus the vertical motions are not particularly close to resonance.  The primary effect is a small time lag in the minima of the scale height with respect to crests in gas column.  The gas frequency
indicates the time-scale $\approx \frac{1}{\Omega_g}$ $\sim$ 10 Myr for the scale height to respond to a change in conditions.  

In the case of present-day disk galaxies the density waves are strongly decoupled from the gas response because the waves are dominated by the old stellar disk which is typically an order of magnitude more massive than the gas disk.  This will change the characteristic frequencies of both the waves and vertical motions so that a more resonant response may be possible.

The local effective support can change substantially with strong stellar feedback in response to star formation.  Feedback energy in any form increases the effective sound speed and causes the scale height to jump in response.  This is particularly apparent in the middle panels of Figure~\ref{fig:response}, corresponding to 100\% feedback efficiency and a reasonable star formation rate.  In this case the pressure terms get substantially out of balance, leading to strong oscillations in the scale height.  The ability of the dynamic pressure model to explain the behaviour of our simulated galaxy disks is explored in the next section.

\subsection{Model with star formation}

The non-equilibrium model from the previous section captures large scale vertical motions of the disk well, but small scale physics such as star formation is a challenge.  To compare to the simulations we make three simple additions to the model.  Firstly, we apply some modest damping to the vertical oscillation on a time-scale comparable to the gas crossing time to mimic shocks and other kinetic energy losses,
\begin{align}
  \frac{\partial \bar{v}_z}{\partial t} &= \frac{2}{\sigg}  \left( \mathrm{P}_{\mathrm{S}}-\mathrm{P}_{\mathrm{R}} \right) - \bar{v}_z/\tau_{\text{D}},
\end{align}
where $\tau_{\text{D}}$ was set to 20 Myr for all the results shown here.   Factors of two variation in this parameter moderately change the range of the scale height variation but do not strongly affect the star formation rate, for example.

Secondly, we need to model star formation.  We elected to use a Schmidt Law similar to that employed in the simulations.  The star formation surface density is thus,
\begin{align}
   \dot{\Sigma}_* = c_* f_{\text{DENS}} {\sqrt{G\,\rho_g}} \sigg.
   \label{eqn:modelsfr}
\end{align}
The density is estimated from the gas column $\rho_g = \frac{1}{2} \sigg/H_g$.  The primary change over the simulation sub-grid model is a correction to estimate the fraction of the gas which is in cold clouds and is thus eligible for star formation.  We used a fixed factor of $f_{\text{DENS}} = 0.1$ for all cases.  It is generally difficult to estimate the detailed density structure.  This limits our ability to use this model to study the simulations where we varied the star formation density threshold.
The parameter $c_*$ is effectively equivalent to that used in the simulations.

Finally, we need to include stellar feedback.  The simple feedback model employed in the simulation readily lends itself to inclusion in a simple model.  Just as in the simulations, $10^{49}$ ergs per solar mass of new stars is applied over a period of 40 Myr and removed on a time-scale of $\tau=5$ Myr.  The energy per unit area, $E_{fb}$, changes as follows,
\begin{align}
\dot{E}_{fb} = \epsilon_{\text{FB}} 10^{49} \mathrm{erg\,M_{\odot}^{-1}}\ \dot{\Sigma}_* - \frac{E_{fb}}{\tau}. \label{eqn:simplefb}
  \end{align}
The parameter, $\epsilon_{\text{FB}}$ is identical to that used in the simulations.  The mid-plane internal energy is given by $u_{fb} = \frac{1}{2} E_{fb}/H_g/\rho_g$.

To summarize, in our time dependent model, effective pressure support can come from two components.  The first is a fixed contribution, that could be attributed to thermal and turbulence from large-scale shear.  The second pressure component varies with the amount of star formation and encompasses supernovae and radiative feedbacks.  In actuality, turbulence arises from many sources and can be strongly correlated with feedback.  However, this is very difficult to characterize within a simple model.  Turbulence could be modelled as a form of feedback with a locally variable characteristic decay time-scale.  In this simple model we use a constant decay time.  The non-cooling feedback energy is open to interpretation and could include turbulence.  

\subsection{Model results}

In Figure~\ref{fig:toymodel} we show our results for the local model with three parameter choices: a high feedback case (corresponding to our reference parameter set), a low feedback case (corresponding to n100.c6.FB10) and a low star formation efficiency case (corresponding to n100.c1.5).  This can be compared directly with Figure~\ref{fig:response}.  The results for the full parameter set are summarized in Table~\ref{tab:sfravg}.  We also include averages from the simulations in Table~\ref{tab:sfravg}.  We stress that the relative changes in the model SFRs versus the simulation SFRs should be compared, rather than the specific values. 

Note that in creating the three model cases, the \textit{only} parameters changed in the model were the feedback and star formation efficiency.  The similarity with Figure~\ref{fig:response} is striking. The local model exhibits the same time variation seen in the simulations.  Vertical motions are driven by pressure imbalances.   The slight lag between star formation peaks and pressure increases that is difficult to distinguish from noise in three-dimensional simulations is now far more apparent.

The match is not perfect, as should be expected, given the simple local model.   The star formation model was kept very basic, avoiding the temptation of introducing tuneable parameters.  A more complex model could be developed but is not necessary to explain the behaviour.  This model works best for large scale effects such as feedback (e.g. varying $\epsilon_{\text{fb}}$).  The inherent small-scale nature of star formation makes it harder to model in the basic framework, particularly non-linear parameters such as thresholds for star formation.  For this reason, we did not attempt to introduce star formation thresholds into the local model.  As a result, the star formation varies more smoothly in the model than it does in the three-dimensional simulations.

The effect of changing the feedback efficiency is shown in the middle- and left-hand panels of Figure~\ref{fig:toymodel}.   A change in feedback efficiency by a factor of 10 increased the star formation by a factor of $\sim$2.5 in the entire simulated disk.  At a radius of 4.5 kpc, as shown in Figure \ref{fig:response}, that difference is  $\sim$ 1.8 times. In the simple model, a similar sub-linear scaling emerges: the same change in parameters produces a change in star formation of $\sim$1.5 times.

The local model is more sensitive to the star formation efficiency ($c_*$) than are the simulations.  In the simulations, at a radius of 4.5 kpc, decreasing the star formation efficiency by a factor of 4 decreases the amount of star formation by $\sim$ 1.6 times.  As shown in Figure~\ref{fig:toymodel} (and Table \ref{tab:sfravg}) decreasing the star formation efficiency by a factor of 4 in the model decreases the star formation rate by $\sim$3.1 times.  This is a stronger scaling than we see in our actual simulations (closer to linear).   As noted earlier, star formation rates depend on ISM properties and small scale structure in a highly non-linear manner that is hard to mimic without resorting to a more complex model.  This can be seen in the more peaked behaviour of the simulation star formation rates in Figure~\ref{fig:response}.  Despite these issues the qualitative behaviour is quite similar.

The simple model is able to reproduce the relative relationships between different pressure support and requirement terms.  For our 100\% feedback efficiency case we saw an equal contribution to the required pressure from either the dark matter of gas component.  The pressure requirement then became dominated by the gas term at lower feedback efficiency (10\%).  Similarly, at high efficiency the largest support came from the pressure due to feedback, with an increased role for the fixed component (which represents thermal and turbulent energy) as the efficiency is decreased.  

We have over-plotted the expected star formation rate from the Kennicutt-Schmidt relation in the top panels of Figure~\ref{fig:toymodel} for reference.  Just as for the simulations, the model star formation rates are lower than this, which we attribute to the lack of an old stellar disk.   As the relation only depends on gas surface density, it is the same in each case.  The different model parameters lead to different outcomes, however.  Even in this simple model, there is plenty of freedom to adjust the parameters to improve the agreement and this calibration process is a common step for most simulation codes.  However, most codes employ even more complex models and there is no guarantee that achieving a match to the relation creates a more realistic simulation overall.  This is particularly true given that most codes are calibrated to isolated models.

The observed Kennicutt-Schmidt relation has considerable vertical scatter.  Models such as this can be used to explore the origins of this.   Both the vertical structure and the star formation rate depend on many factors, such as the stellar surface density, the rotation curve, the two-phase nature of the ISM and so forth.  This was explored for local galaxies by \cite{ostriker10}.

\subsection{The origin of sub-linear scaling}\label{sublinear}

For the cases shown in Figure~\ref{fig:toymodel} here we have assumed an oscillating surface density.  Using a constant surface density decreases the average star formation rate and scale height by $\sim$10\%.  Holding the surface density constant causes the height oscillations to damp away.  We reach an equilibrium where the required pressure equals the supporting pressure.  This is equivalent to the equilibrium model of \cite{ostriker10}.  Thus in terms of averaged quantities, an equilibrium model is sufficient.

Setting $P_{\text{R}} - P_{\text{S}} = 0$ as in the right-hand side of equation~\ref{eqn:dh2dt2} and keeping only those terms relevant to these simulations gives
\begin{align}
\frac{1}{H_g} (\gamma-1) (u_{\mathrm{fb}} + u_{\mathrm{other}})
- \Omega^2 H_g  - \pi G \sigg  = 0. \label{eqn:ugly}
\end{align}
With our assumed Schmidt law (equation~\ref{eqn:modelsfr}) and the simple feedback model (equation~\ref{eqn:simplefb}), the equilibrium feedback energy is,
\begin{align}
  u_{\text{fb}} = \frac{\sqrt{G}}{2}\,\tau\, c_*\, f_{\text{DENS}}\,\epsilon_{\text{fb}}\,(10^{49} \mathrm{erg\,M_{\odot}^{-1}})\ \sigg^{1/2}\, H_g^{-1/2}. \label{eqn:ufb}
\end{align}
When this is substituted into equation~\ref{eqn:ugly}, the result is a non-linear equation for $H_g$.  The solution for the mean scale height is within 10\% of the average of $H_g$ in figure~\ref{fig:response}.
As is clear from equation~\ref{eqn:ufb}, $c_*$ and $\epsilon_{\text{fb}}$ affect the scale height in the same way.  The is exactly true in the local model and approximately true in the full simulations.  The non-linear dependence of $H_g$ on these quantities is the reason for the non-linear scaling seen in the simulations.

 For parameter values in the region of interest, the scale height depends on $c_*$ and $\epsilon_{\text{fb}}$ roughly as $H_g \sim c_*^{0.4} \epsilon_{\text{fb}}^{0.4}$.  The star formation rate differs from the energy expression by one power in $c_*$ and changes as $H_g^{-1/2}$.  Thus the star formation rate depends on these parameters roughly as $\dot{\Sigma}_*\,\sim\,c_*^{0.8}\ \epsilon_{\text{fb}}^{-0.2}$.  This is essentially what we see in the local model.  These particular power-laws are specific to this radius and local model parameters.  However, similar sub-linear behaviour will occur generally.

\section{Summary and Conclusions}
\label{sec:conclusions}
We have presented a suite of high resolution isolated galaxies that include a purposefully simple feedback scheme.  We use a wide range of parameter choices to explore both the implications of these choices and the process of regulation of star formation.  We find a sub-linear scaling between parameter choices and the resulting amount of star formation.  This is found in many other simulations \citep[see, e.g.][]{hopkinsQM11}.   We adapted the equilibrium pressure regulation models of \cite{ostriker10} and \cite{kimKimOstriker11, kimOstrikerKim13} to include representative simulation sub-grid recipes for star formation and feedback.  This allowed us to demonstrate the origin of the sub-linear scaling with these parameters.  The equilibrium models readily explain how feedback affects the star formation rates.  The effect of changing the overall star formation efficiency is also well represented by the models, however, strongly non-linear parameters, such as star formation density thresholds, are harder to incorporate.  Such parameters also reflect the complexity of star-formation and how difficult it is to model even in three-dimensional simulations.

  The simulations show regular variations in the local star formation rates and other properties.  These are driven by density waves which occur naturally in global models.  In comparison, small box simulations \citep[e.g.][]{kimOstrikerKim13} are expected to approach rough equilibrium and do not need to consider vertical oscillations.   We extended the equilibrium models into a dynamic pressure-driven regulation framework.  We show that advection plays a minor role and thus the vertical motions are driven by local effective pressure differences.  We adapt the star formation and feedback models used in the simulations to produce a complete local model of star formation and feedback.  These models are able to qualitatively reproduce the behaviour of our simulated galaxies, including the variability and the sub-linear scaling with the star formation and feedback efficiency parameters.

  A goal of this work was to determine which aspects of the small-scale star formation physics strongly affect larger scale simulations.  The simulation community has invested in a diverse array of feedback prescriptions and types.  We demonstrate that the crucial factor is how feedback translates into effective pressure support on larger scales.  It is the effective pressure support that regulates star formation and the vertical structure of the ISM.  Realistic regulated star formation thus requires that the scale height be resolved.

Variable FUV backgrounds are a potentially important stellar feedback for normal spiral galaxies.  Including this variability is feasible in local boxes \citep{kimOstrikerKim13} but is numerically challenging in global galaxy simulations.  In future work, we plan to explore this mode of feedback using newly developed radiative transfer techniques (Woods et al., in preparation) including an old stellar disk.  The inclusion of FUV that is coupled to star formation allows self-consistent measurement of the partition between different pressure components \citep{kimOstriker15,koda16}.

\section*{Acknowledgements}
The authors thank the referee, Chang-Goo Kim, for a detailed report that greatly improved the paper.  
The authors thank Elizabeth Tasker, Alex Pettitt and Diederik Kruijssen for helpful discussions.    The authors thank SHARCNET (Shared Hierarchical Academic Research Computing Network), SciNet, and Compute/Calcul Canada, who provided dedicated resources to run these simulations.  Computations were performed on the gpc supercomputer at the SciNet HPC Consortium.  SciNet is funded by: the Canada Foundation for Innovation under the auspices of Compute Canada; the Government of Ontario; Ontario Research Fund - Research Excellence; and the University of Toronto.  Data visualization was performed using the yt\footnote{http://yt-project.org} toolkit by \cite{turk11}.  This work was supported by NSERC.  SMB acknowledges financial support from the Vanier Canada Graduate Scholarship program.

\bibliographystyle{mnras}
\bibliography{regulation_report2}

\appendix
\section{Pressure Terms}
\label{app:pressuremath}

In the pressure driven regulation framework of \cite{ostriker10}, gas weight set
through vertical gravity creates a pressure requirement in the ISM that must be
matched by some type of pressure support.  We begin by examining the components of the galaxy that contribute to vertical gravity, namely dark matter, gas and stars.  These add linearly to give a required mid-plane pressure,
\begin{equation}
 P = \int^{\infty}_0 \rho g\, \text{d}z, \notag
\end{equation}
where $\rho_g$ is the gas density and $g$ is the total gravity.  The
contribution to the gravity from dark matter, which is taken here to include all
spherically symmetric large-scale components of the galaxy, is directly related
to galactic rotation and can be estimated near the disk as,
\begin{equation}
 g_{\text{dm}} = \Omega^2 z,
\end{equation}
where $\Omega$ is the angular rotation rate.  Then the pressure required for support against gravity from the dark matter component is
\begin{align}
  P_{\text{dm}} &= \int^{\infty}_0 \rho_g g_{\text{dm}}\, \text{d}z  \\
  &= \frac{1}{2} \Omega^2 H_g \sigg \label{eqn:Pdm},
\end{align}
where $H_g$ and $\Sigma_g$ are the scale height and surface density of the gas, respectively.

Gravitational acceleration due to planar, disk material is given by,
\begin{equation}
 g = 2\pi G \Sigma (z),
\end{equation}
where $ \Sigma (z)$ is the disk surface density within a height, $z$, of the mid-plane.  Thus the pressure required for support against self-gravity from the gas in the disk is,
\begin{align}
 P_{\text{g}} &= \int^{\infty}_0 \rho_g g_{\text{g}}\, \text{d}z \\
  &= \frac{1}{2} \pi G \sigg^2 \label{eqn:Pgas},
\end{align}
where now $ \sigg$ denotes the total gas surface density.

Similarly, the expression for stellar gravity is,
\begin{equation}
 g_* = 2\pi G \Sigma_* (z),
\end{equation}
where $ \sigg (z)$ is the gas surface density at a given height above the mid-plane.  The pressure required for support against stellar gravity is, 
\begin{align}
P_* &= \int^{\infty}_0 \rho_g g_*\, \text{d}z.
\end{align}

\begin{figure*}
\centering
\begin{tabular}{cc}
\includegraphics[scale=0.55]{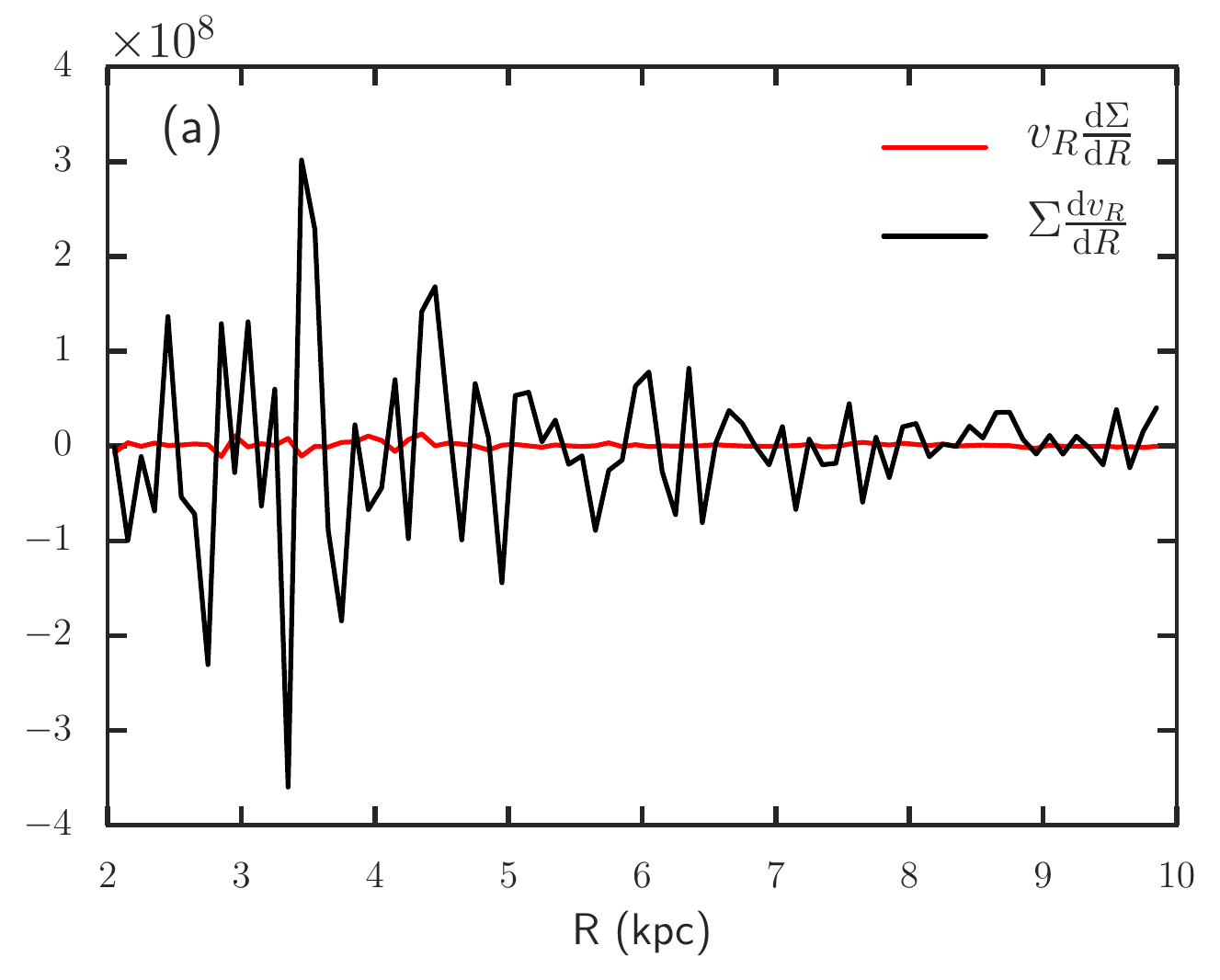} &
\includegraphics[scale=0.55]{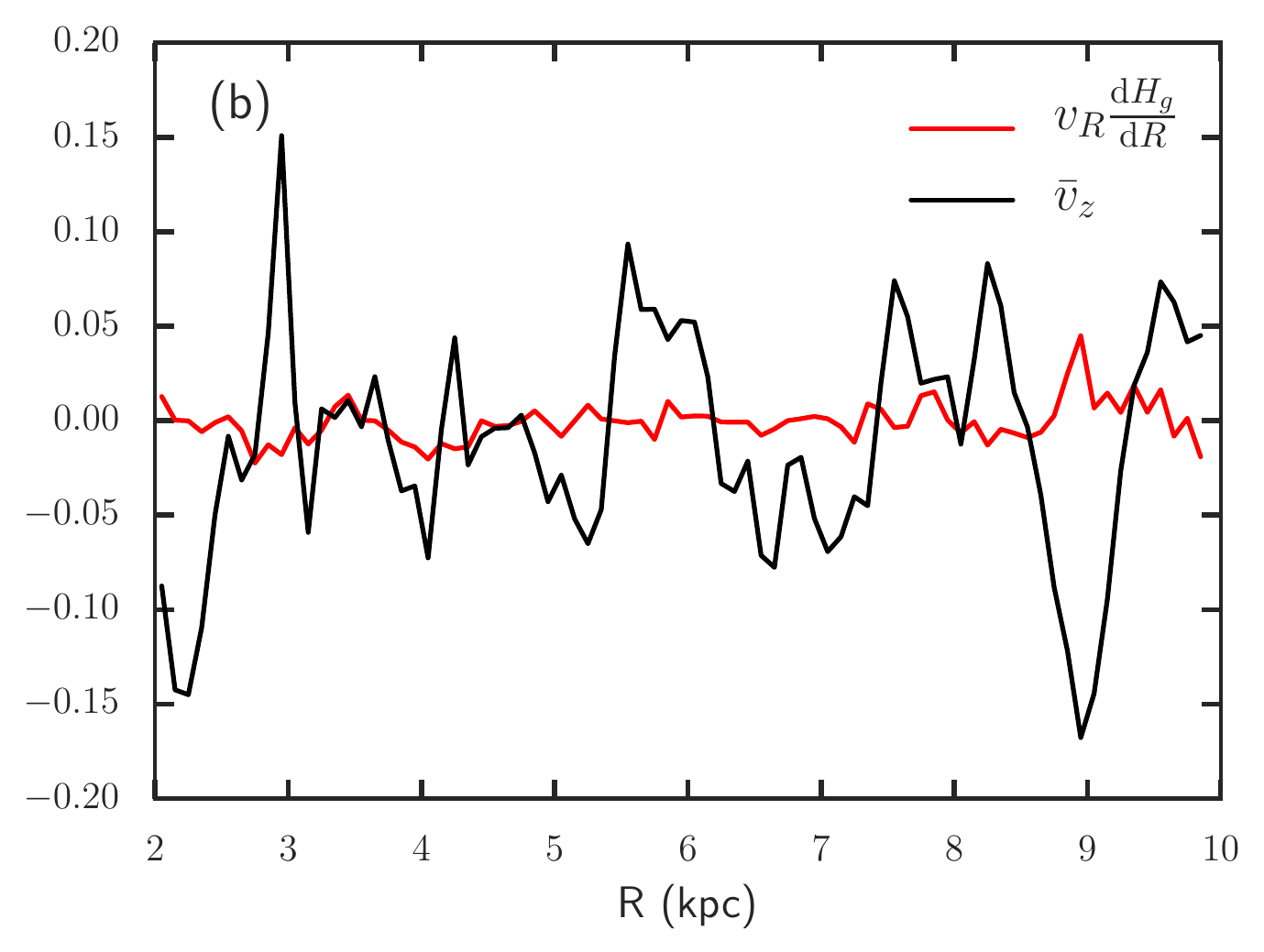}\\ 
\end{tabular}
\caption{{\it Left:} Planar advection vs. compression term in the rate of change of gas column.  {\it Right:} Planar advection vs. vertical motion term in the rate of change of scale height.  In both cases the contribution of advection in the plane is small.}
\label{sigmanoadvect}
\end{figure*}

A solution for the pressure needed to balance the stellar component requires an assumption for the functional form of the gas and stellar densities.   We assume the following form,
\begin{align}
 \rho_g &= \rho_{g,0} \text{e}^{-z/H_g}\\
 \rho_* &= \rho_{*,0} \text{e}^{-z/H_*}.
\end{align}
Then an expression for the stellar surface density is
\begin{align}
 \Sigma_*(z) &= 2 \int^z_0 \rho_* \text{d}z  \\
  &= \Sigma_*\left(1 - \text{e}^{-z/H_*}\right),
\end{align}
where $H_*$ is the stellar scale height, and $\rho_{g,0}$ and $\rho_{*,0}$ are the initial gas and stellar densities, respectively.  Finally, the required pressure support for the stellar component is
\begin{align}
 P_* &= 2 \pi G\int^{\infty}_0 \rho_g \Sigma_* (z)\, \text{d}z\\
 &= 2 \pi G\Sigma_*\rho_{g,0} \int^{\infty}_0 \text{e}^{-z/H_g} - \text{e}^{-z\left(\frac{1}{H_g} + \frac{1}{H_*}\right)} \text{d}z\notag\\
  &= 2 \pi G\Sigma_*\rho_{g,0} H_g \left[1 - \left(\frac{H_g}{H_g} + \frac{H_g}{H_*}\right)^{-1}\right]\notag\\
 P_*&= \pi G\Sigma_*\Sigma_g \left(\frac{H_g}{H_g+H_*}\right),
 \label{eqn:Pstar}
\end{align}
If we assume the stellar scale height $H_*$ is small, as is appropriate for a very young stellar population, we get an upper bound on the stellar contribution to the total pressure,  $ P_* = \pi G\Sigma_*\Sigma_g$.  As shown in section~\ref{sec:balance}, the net contribution of stellar gravity in initially star-free, isolated disk tests such as the ones employed here is small.

In the case that the scale height of the stars is large, as in the case of an old stellar disk, the surface density associated with stars approaches a linear function of height, $\Sigma_* \approx 2 \rho_{*,0}\,z$ as used in \cite{ostriker10}, where $\rho_{*,0}$ is the mid-plane stellar density.  Then the required pressure is approximately $P_* \approx 2 \pi G \Sigma_g H_g \rho_{*,0}$ which matches \ref{eqn:Pstar} for the case when $H_g \ll H_*$.

\section{Dynamic Pressure Balance}
\label{app:presmodel}

Assuming symmetry about the $z=0$ plane, the scale height of the gas may be defined:
\begin{align}
  H_g &= \frac{\int^{\infty}_0 \rho_g z\ \text{d}z}{\int^{\infty}_0 \rho_g\ \text{d}z}\nonumber \\
    &= \frac{2}{\sigg} {\int^{\infty}_0 \rho_g z\ \text{d}z},
\end{align}
The scale height changes in response to changes in conditions, in particular the gas column, ${\sigg}$.  The change in the gas column with time is dominated by density waves moving solely in the plane of the disk,
\begin{align}
  \frac{\partial \sigg}{\partial t} &= -\vec{v}_{R\phi} \cdot \nabla \sigg - \sigg \nabla \cdot \vec{v}_{R\phi},
  \label{eqn:dsigdt}
\end{align}
where $\vec{v}_{R\phi}$ denotes motions in the plane that are assumed to be independent of $z$.  This allows us to take planar velocity terms out of the integrals where necessary.

For linear waves, the advection term (first on the right in equation \ref{eqn:dsigdt}) is second order and negligible compared to the second term associated with compressive waves.  In a non-linear disk scenario, fractional variations in $\sigg$ can be order unity, but the evolution of the gas column is still dominated by compressive rather than material waves and individual gas elements travel through the wave crests rather than with them.  This is demonstrated to be the case for the simulations used here in Figure \ref{sigmanoadvect}(a).

$H_g$ changes over time due to the movement of gas both in the plane and vertically,
\begin{align}
  \frac{\partial H_g}{\partial t} &= -\frac{H_g}{\sigg} \frac{\partial \sigg}{\partial t}  + \frac{2}{\sigg} \int^{\infty}_0 \frac{\partial (\rho_g z)}{\partial t}\ \text{d}z.
\end{align}
The second term can rewritten using the conservation of mass,
\begin{align}
\frac{2}{\sigg} \int^{\infty}_0 \frac{\partial (\rho_g z)}{\partial t}\ \text{d}z   = \frac{2}{\sigg}   \int^{\infty}_0 \frac{\partial \rho_g}{\partial t}\ z\ \text{d}z\nonumber \\
  = -\frac{2}{\sigg}   \int^{\infty}_0 \left( \nabla \cdot (\rho_g \vec{v}_{R\phi}) + \frac{\partial (\rho_g v_z)}{\partial z} \right)\ z\ \text{d}z, \label{eqn:consmass}
\end{align}
where the gradient, $\nabla$, is used here to represent the gradient in the two planar directions ($R$ and $\phi$) only.
The divergence term in the planar velocity can be broken into two parts,
\begin{align}
  & \frac{2}{\sigg}  \int^{\infty}_0 \nabla \cdot (\rho_g \vec{v}_{R\phi})\ z\ \text{d}z \notag \\
& = \frac{2}{\sigg} \int^{\infty}_0  \vec{v}_{R\phi} \cdot \nabla (\rho_g z)\ \text{d}z +
  \frac{2}{\sigg} \int^{\infty}_0  ( \rho_g z ) \nabla \cdot \vec{v}_{R\phi} \ \text{d}z \nonumber \\ 
  & = \vec{v}_{R\phi} \cdot \nabla H_g
  + \frac{H_g}{\sigg} \vec{v}_{R\phi} \cdot \nabla \sigg
  + \frac{2\,H_g}{\sigg} \nabla \cdot \vec{v}_{R\phi} \notag \\
  & = \vec{v}_{R\phi} \cdot \nabla H_g - \frac{H_g}{\sigg} \frac{\partial \sigg}{\partial t},
\end{align}
where the last line makes use of equation~\ref{eqn:dsigdt}.

The vertical mass flux term from \ref{eqn:consmass} simplifies as follows,
\begin{align}
   \frac{2}{\sigg}  \int^{\infty}_0 -\frac{\partial (\rho_g v_z)}{\partial z}\ z\ \text{d}z
  &= \frac{2}{\sigg}  \int^{\infty}_0 \rho_g v_z\ \text{d}z = \bar{v}_z, 
\end{align}
using integration by parts and symmetry about the $z=0$ plane.  This term is the mass weighted mean vertical velocity in the half-plane, which we will denote by, $\bar{v}_z$. 

Combining terms, we get a simple planar advection term for $H_g$ plus changes due to vertical motions in the form of $\bar{v}_z$.  Thus the comoving rate of change of the scale height is,
\begin{align}
\frac{d H_g}{d t} &= \frac{\partial H_g}{\partial t} + \vec{v}_{R\phi} \cdot \nabla H_g = \bar{v}_z .\label{eqn:dhdt}
\end{align}
As $H_g$ is a ratio of density-weighted integrals, there is no equivalent of the compressive term for the gas column.
As with the gas column, we expect the advection term to play a minor role so that changes in $H_g$ are dominated by the mean vertical velocity.  This is precisely true in the frame of fluid elements moving within the disk and passing through density waves.
We show this is true at fixed points in our simulations in figure~\ref{sigmanoadvect}(b), confirming the unimportance of advection terms.

The rate of change of the mean vertical velocity is given by,
\begin{align}
  \frac{\partial \bar{v}_z}{\partial t} &= -\frac{\bar{v}_z}{\sigg} \frac{\partial \sigg}{\partial t} + \frac{2}{\sigg} \int^{\infty}_0 \frac{\partial (\rho_g v_z)}{\partial t}\ \text{d}z.
\end{align}
Using the momentum conservation equation we can write,
\begin{align}
& \frac{2}{\sigg} \int^{\infty}_0 \frac{\partial (\rho_g v_z)}{\partial t}\ \text{d}z
  \notag \\
  & =  \frac{2}{\sigg}   \int^{\infty}_0 \left( -\nabla \cdot (\rho_g \vec{v}_{R\phi}) -\frac{\partial (\rho_g v_z^2)}{\partial z} -\frac{\partial \mathrm{P}}{\partial z} + \rho_g g \right)\ \text{d}z,
\end{align}
where $P(z)$ denotes the pressure and $g(z)$ is the vertical gravitational acceleration.  Just as for the scale height, the sole remaining term from the $\partial \sigg/\partial t$ terms and the planar velocity terms is a term for the advection of $\bar{v}_z$.  As a result we can express the comoving rate of change of the vertical velocity as,
\begin{align}
  \frac{d \bar{v}_z}{d t}  &=   \frac{\partial \bar{v}_z}{\partial t} +
  \vec{v}_{R\phi} \cdot \nabla \bar{v}_z \notag \\
  &= \frac{2}{\sigg}    \left( \left. \rho_g v_z^2 \right|^{\infty}_0 - \left. \mathrm{P} \right|^{\infty}_0 + \int^{\infty}_0 \rho_g g\ \text{d}z \right) \nonumber \\
  &= \frac{2}{\sigg}  \left( \mathrm{P}_{\mathrm{S}}-\mathrm{P}_{\mathrm{R}} \right). \label{dvzdt}
\end{align}
Thus the change in the mean vertical velocity depends on the difference between the vertically integrated weight or required pressure, $\mathrm{P}_{\mathrm{R}}$ and the mid-plane supporting pressure, $\mathrm{P}_{\mathrm{S}}$.  These pressures were defined in section~\ref{sec:balance} and their individual components calculated in appendix~\ref{app:pressuremath}. These equations describe the dynamic pressure balance model.

As discussed above, planar advection is unimportant for density waves in the disk.  Thus we can use the comoving and partial time derivative interchangeably for $H_g$ and $\bar{v}_z$.

\bsp	
\label{lastpage}
\end{document}